%% file: main.tex
\documentclass[conference,compsoc]{IEEEtran}

\ifCLASSOPTIONcompsoc
  \usepackage[nocompress]{cite}
\else
  \usepackage{cite}
\fi

\ifCLASSINFOpdf

\else

\fi

\usepackage{amsthm}
\usepackage{array}
\usepackage{amsmath,amssymb,amsfonts}
\usepackage{caption}
\usepackage{algorithm}
\usepackage{algpseudocode}
\usepackage{multirow}
\usepackage{multicol} 
\usepackage{graphicx}
\usepackage{epsfig}
\usepackage{cite}
\usepackage[labelfont=bf,labelsep=quad]{caption}
\usepackage{booktabs}
\usepackage{bm}
\usepackage{graphicx}
\usepackage{subfig}
\usepackage[top=1.425cm, bottom=1.425cm, left=1.425cm, right=1.425cm]{geometry}
\usepackage{color}
\usepackage{xcolor}    
\usepackage{diagbox}
\usepackage{makecell}
\usepackage{xspace}
\usepackage{textcomp}
\usepackage{arydshln}
\usepackage{booktabs}
\usepackage{supertabular}
\usepackage{setspace}
\usepackage{float}
\usepackage{epstopdf}
\usepackage{mathtools}
\usepackage{amsthm}
\usepackage{tabularx}
\usepackage[hidelinks]{hyperref}
\usepackage{threeparttable}
\newtheorem{lemma}{Lemma}

\def\eg{\textit{e.g.}\xspace}
\def\ie{\textit{i.e.}}

\def\sdae{DFSD\xspace}
\def\ane{FNE\xspace}

\def\usi{ID\xspace}
\newcommand{\name}{\emph{TAAC}\xspace}

\begin{document}
\title{\name: A gate into Trustable Audio Affective Computing}

\author{
\IEEEauthorblockN{Xintao Hu}
\IEEEauthorblockA{
\textit{Hefei University of Technology}\\
Hefei, China}

\and

\IEEEauthorblockN{Feng-Qi Cui}
\IEEEauthorblockA{
\textit{Hefei University of Technology}\\
Hefei, China}}

\maketitle

\input{chapters/chapter1-Abstract}


\IEEEpeerreviewmaketitle

\input{chapters/chapter2-Introduction}
\input{chapters/chapter3-RelatedWork}
\input{chapters/chapter4-PreliminariesAndObservation}
\input{chapters/chapter5-SystemDesignandMethodology}
\input{chapters/chapter6-ExperimentalEvaluation}

\input{chapters/chapter7-Conclusion}
\input{chapters/Chapter-9-EthicConsideration}

\ifCLASSOPTIONcompsoc
  \section*{Acknowledgments}
\else
  \section*{Acknowledgment}
\fi

Anonymous for now.

\bibliographystyle{IEEEbib}
\bibliography{bib/KeyStroke.bib}

\end{document}

%% file: chapters/chapter1-Abstract.tex
\begin{abstract}

With the emergence of AI techniques for depression diagnosis, the conflict between high demand and limited supply for depression screening has been significantly alleviated. Among various modal data, audio-based depression diagnosis has received increasing attention from both academia and industry since audio is the most common carrier of emotion transmission. Unfortunately, audio data also contains User-sensitive Identity Information (\usi), which is extremely vulnerable and may be maliciously used during the smart diagnosis process. Among previous methods, the clarification between depression features and sensitive features has always serve as a barrier. It is also critical to the problem for introducing a safe encryption methodology that only encrypts the sensitive features and a powerful classifier that can correctly diagnose the depression.
To track these challenges, by leveraging adversarial loss-based Subspace Decomposition, we propose a first practical framework \name presented for Trustable Audio Affective Computing, to perform automated depression detection through audio within a trustable environment. 
The key enablers of \name are Differentiating Features Subspace Decompositor (\sdae), Flexible Noise Encryptor (\ane) and Staged Training Paradigm, used for decomposition, ID encryption and performance enhancement, respectively. 
Specifically, to separate different types of features for protecting user privacy, by employing subspace-based distenglement loss to dual-end Auto Encoder, we propose \sdae to maximally decompose the features between id-related and depression-related. 
Then, by employing progressive noise addition, we propose a Deterministic Diffusion (Encryption) process and introduce \ane, for encrypting the id-related data with adjustable encryption strength.
Then, by goal-oriented mulfunctional segment-based collaborative training, a noval staged training paradigm that enables different modules to specialize in their respective predefined functions is designed.
Extensive experiments with existing encryption methods demonstrate our framework's preeminent performance in depression detection, \usi reservation and audio reconstruction. Meanwhile, the experiments across various setting demonstrates our model's stability under different encryption strengths. Thus proving our framework's excellence in Confidentiality, Accuracy, Traceability, and Adjustability.

\end{abstract}

\begin{IEEEkeywords}
Depression Diagnosis, Personal Information Protection, Security, Trustable Affective Computing
\end{IEEEkeywords}

%% file: chapters/chapter2-Introduction.tex
\section{Introduction}
\subsection{Backgrounds and Motivations}


Depression is a widespread mental health condition affecting an estimated 5\% of adults globally \cite{WHO2023Depression, WHO2025Depression}. Left untreated, depression can escalate to more severe symptoms. Recently, rapid and efficient audio-based AI solutions, such as automated PHQ-9 screenings \cite{PHQ-9-1}, have been integrated into clinical workflows, which has greatly helped mitigate the shortage of mental health professionals.

Compared to visual-based depression detection methods, audio-driven approaches offer enhanced convenience and stronger privacy preservation for users. However, \textbf{Trustability} remains underexplored in depression detection systems across both visual and acoustic modalities. Academic studies have long emphasized that audio contains voiceprints embedding User-sensitive Identity Information (hereafter refer as \usi) \cite{VoicePrint-1}. Recent advances in speaker verification, such as the SpeechBrain framework achieving a 0.69\% Equal Error Rate \cite{ravanelli2021speechbrain}, demonstrate the near-biometric precision of current \usi extraction technologies. More critically, during PHQ-9 assessments, users are required to verbally respond to sensitive questions about personal experiences and emotions, the leakage of such voice recordings could enable large-scale \usi breaches.

In real-world deployments, while patients provide raw audio recordings containing embedded \usi, they lack technical mechanisms to audit or constrain downstream data usage \cite{USENIX-audio-1}, \cite{SP-audio-1}. Malicious actors could exploit these vulnerabilities through multiple attack vectors, such as bulk-reselling speech samples on black or reconstructing identity vectors from audio snippets to facilitate voice phishing campaigns. A particularly concerning scenario involves critical institutional processes, such as mandatory depression screenings during corporate hiring. If audio data collected during corporate mental health assessments is compromised, employee voiceprints - directly linked to their identities, professional roles, and organizational occupations, could be illicitly acquired by third-party headhunters. 

\begin{figure}
    \centering
    \includegraphics[width=1\linewidth]{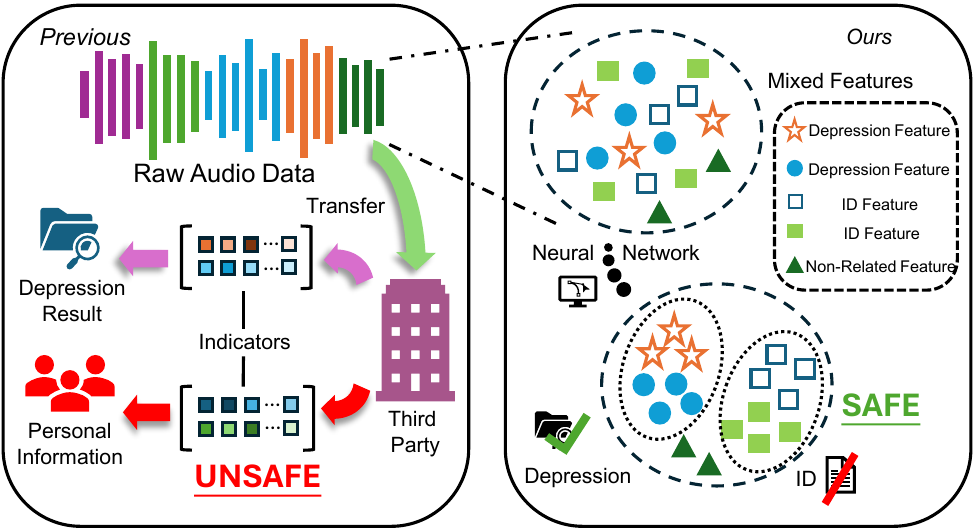}
    \caption{The kernel insight of our implementation}
    \label{fig:img3}
\end{figure}


Early research efforts, in audio-based depression detection have primarily focused on improving diagnostic accuracy, often assuming raw audio data is explicitly accessible and trustworthy. Many studies leverage semantic content (\eg, spoken words) extracted directly from audio signals to infer depression severity \cite{ye2021multi, shen2022automatic, CCS-1}, which, under trustworthy AI frameworks, represent sensitive attributes requiring protection.

\subsection{Challenges and Contributions}
Three major chanllenges need to be formally addressed before establishing a trustable audio depression detection environment (Hereafter, the data owner is referred to as Party A, and the delegated processor conducting depression detection is termed Party B):

\begin{itemize}
    \item \textbf{Confidentiality}: In a trustable environment, mechanisms must be established to prevent Party B from directly accessing \usi embedded in the audio data. This necessitates supplementing additional obfuscation methods to mask identity-related features. Crucially, Party B must be technically restricted from correlating audio samples with their speakers.

    \item \textbf{Accuracy}: While obfuscating \usi in audio data, we must ensure that depression-related information can still be extracted reliably and that the system accurately detects whether a speaker has depression—without ever learning the speaker's identity.

    \item \textbf{Traceability}: While ensuring Party B cannot access identity information in the audio data, Party A must retain the capability to reconstruct the original audio and map depression detection results back to individual speakers upon receiving processed data from Party B.
\end{itemize}

An additional requirement may be introduced that Party A should be able to dynamically adjust the encryption strength based on operational priorities, lowering privacy-preserving intensity in scenarios prioritizing detection accuracy while increasing it in contexts demanding heightened security. This flexibility inherently allows data owners to make context-aware tradeoffs between analytical fidelity and confidentiality, ensuring compliance with both clinical and organizational constraints. 

To address these challenges, this paper proposes a novel privacy-preserving depression detection framework. As illustrated by \ref{fig:img3}, in the lagacy framework, Party B is capable of illegally misusing identity information. Whereas in the proposed framework, Party B is UNABLE to extract the identity information while Party A's ability to reconstruct identity attributes is preserved.

However, within this novel framework, the co-design of encryption modules and feature extraction networks remains a critical challenge, requiring these components to be carefully architected to achieve two objectives: (1) preserving sufficient information for accurate depression detection, and (2) guaranteeing the computational irreversibility of \usi. To achieve these objectives, we design three major components that integrates Subspace Decomposition for feature disentanglement and Noise-diffusion Encryption for identity obfuscation, which employs hierarchical progressive training with multiple losses to isolate depression biomarkers from speaker-specific attributes. Implementation details will be introduced in Sec. \ref{sec:SysDes}.

In brief, our contributions can be broadly summarized as follows:
\begin{itemize}
    \item To the best of our knowledge, this is the first work to formalize and address the trustability challenge in depression detection, with a particular focus on data protection in practical deployment.
    \item We propose a Differentiating Features Subspace Decomposition (DFSD) network that employs subspace-based disentanglement loss on a dual-end autoencoder, effectively separating depression-related features from \usi.
   \item To support diverse application scenarios with varying security levels, we develop a Deterministic Tensor Diffusion (DTD) encryption strategy based on conditional probability and progressive noise injection via a Markov chain, achieving a flexible balance between detection accuracy and information confidentiality while resisting advanced decryption models.
   \item We design a goal-oriented, multifunctional, segment-based collaborative training paradigm that enables each module to specialize in its predefined role, improving both depression recognition accuracy and feature disentanglement.
    \item Extensive experiments confirm that the proposed framework achieves complete \usi obfuscation while maintaining near-state-of-the-art accuracy in depression detection.
\end{itemize}

The rest of the paper is organized as follows. Sec. \ref{sec:RelWor} and \ref{sec:PreObs} discusses numerous related studies and preliminaries. Then, we describe the \name system design in Sec. \ref{sec:SysDes}. Implementation, evaluation, and the impacts of various factors on \name performance are presented in Sec. \ref{sec:ExpEva}. Finally, we conclude our work in Sec. \ref{sec:Con}

%% file: chapters/chapter3-RelatedWork.tex
\section{Related Work} \label{sec:RelWor}
\subsection{Depression Detection}
Depression Detection aims to automatically identify depressive symptoms and assess their severity levels by analyzing an individual's behavioral patterns, speech, physiological signals, or neural activity. Its applications include telemedicine, mental health screening, and personalized treatment \cite{NDSS-1, USENIX-2}. In recent years, the integration of multimodal data (such as speech, text, video, and physiological signals) with deep learning techniques has emerged as a major focus of research in this field \cite{SP-2, CCS-2}.

The Audio/Visual Emotion Challenge (commonly known as AVEC) \cite{ringeval2019AVEC, USENIX-3} is an internationally renowned multimodal emotion recognition competition\footnote{AVEC benchmarks ran annually from 2013 to 2019; the first three years (2013-2015) and the final year (2019) are specifically cited here.}. Since 2013, it has repeatedly hosted the Depression Sub-Challenge, a dedicated task for depression detection, significantly advancing algorithmic development in this field. Its primary datasets include DAIC-WOZ \cite{gratch2014DAIC}, D-Vlog \cite{yoon2022Dvlog}, and MODMA \cite{cai2020MODMA}, among others.

In depression detection tasks, many current models have adopted Language Models (LMs) or transformer-based architectures, integrating transformers into hybrid frameworks to leverage their contextual comprehension capabilities for enhanced depression screening \cite{fan2024Transformer, qin2025Language, Yongfeng2024DepMSTAT, SP-2}. Moreover, a growing number of models are adopting multimodal fusion techniques, integrating data across diverse modalities (Audio, Text, Video, Biological Signals) for depression detection, leveraging comprehensive multi-aspect information to improve diagnostic accuracy \cite{NDSS-2, CCS-2}. 

In the domain of audio-based depression detection, prior research has explored various methodologies, Including Graph Neural Networks (GNNs) \cite{Sun2024novel} to model latent connections within/between audio signals and hybrid frameworks or utilizing MFCC-derived spectrogram features processed through Convolutional Neural Networks (CNNs) \cite{das2024deep}. Additionally, some studies have proposed using the pretrained Wav2Vec model for feature extraction, combined with lightweight fine-tuning networks, achieving highly effective detectionn results \cite{huang2024depression}.

The table \ref{tab:sum} lists some of representative methods with their reported classification metrics (Precision, Recall, F-1 score and Accuracy). 

\begin{table*}[t]
\begin{threeparttable}
\caption{A brief summary of representative methods.}
\label{tab:sum} 
\renewcommand\arraystretch{1.2}
\centering
\begin{tabularx}{\linewidth}{lXXXX}
\toprule[1.5pt]
Method & PRE & REC & F1 & ACC \\ 
\midrule[1.5pt]
Unsupervised Autoencoder \cite{sardari2022audio} & 85\% & 80\% & 83\% & - \\  
\midrule
MFCC-based RNN \cite{Emna2022MFCC} & 75\%(70\%) & 95\%(26\%) & 84\%(38\%) & 74\% \\ 
\midrule 
Feature Fusion based on spectral analysis \cite{Xiaolin2022Fusing} & 83\% & 100\% & 91\% & 85\% \\
\midrule
Attentive LSTM Network \cite{yan2021Detecting} & - & - & 94.01\% & 90.2\% \\
\midrule
Graph Neural Network \cite{Sun2024novel} & 92.36±2.53\% & 92.18±1.55\% & 92.23±2.01\% & 92.21±1.86\% \\
\midrule
MFCC \& CNN generated spectrogram features \cite{das2024deep} & 86.4\%(93.4\%) & 93.6\%(85.9\%) & 89.85\%(89.5\%) & 90.26\% \\
\midrule
Voice-based Pre-training Model \cite{huang2024depression} & 93.96\%(97.65\%) & 94.87\%(97.21\%) & 94.41\%(97.43\%) & 96.48\% \\
\bottomrule
\end{tabularx}
\begin{tablenotes} 
\item[*] This table only lists existing methods and their reported results on DAIC-WOZ dataset. Due to differences in experimental setups, these results cannot be directly compared. The strength of our model lies in its ability to process encrypted data, and therefore the accuracy on raw data is not the focus of our investigation.
\end{tablenotes}
\label{table4}
\vspace{0.25in}
\end{threeparttable}
\end{table*}

\subsection{Audio Encryption}
Chaos Maps-Based Encryption (CMBE) is a data protection technique rooted in the properties of nonlinear dynamical systems, which has garnered significant attention in the field of audio encryption in recent years. Primarily applied to audio transmission and storage, this method leverages dynamic response capabilities and resistance to statistical analysis. A defining feature is that even slight variations in the encryption key trigger drastic changes in output, while the generated chaotic sequences mimic random noise, effectively obscuring the statistical characteristics of the original audio signals \cite{albahrani2021review}. 

In addition to chaotic encryption, several other encryption algorithms have been proposed. For example, the Cosine Number Transform (CNT) leverages recursive finite-field transformations and overlapping block selection rules to achieve global diffusion \cite{ustubioglu2022cnt}. Alternatively, Permutation-Substitution Architecture-based methods employ bit-level scrambling operations followed by dynamic substitution driven by pseudo-random sequences \cite{alanazi2023novel}.

Furthermore, homomorphic encryption (HE) is a cryptographic technique frequently employed in cloud computing and machine learning scenarios. It enables direct arithmetic operations (\eg, addition, multiplication) on encrypted data, ensuring that decrypted results match those obtained from equivalent computations performed on plaintext. Its defining property lies in preserving the operational utility of data while remaining encrypted \cite{nguyen2024efficient}.

%% file: chapters/chapter4-PreliminariesAndObservation.tex
\section{Preliminaries And Observations} \label{sec:PreObs}


\subsection{Subspace Decomposition}
At first glance, deploying conventional encryption methods to securely transmit audio data to Party B for depression detection via specialized models seems theoretically feasible. However, our analysis reveals a critical limitation: conventional encryption methods prioritize data confidentiality by transforming raw signals into randomized noise patterns. While this ensures security, the inherent unpredictability and non-decryptability of such encrypted data severely impede the model's ability to identify depression-specific features from chaotic representations, thereby undermining detection accuracy.

Homomorphic Encryption (HE) has also been explored for this task. While HE enables encrypted audio processing within depression detection models \cite{nguyen2024efficient}, empirical evidence demonstrates that despite voice content being obfuscated, \usi extraction models can still reliably determine whether two encrypted audio clips originate from the same speaker, indicating that the individual's \usi remains exposed.

Thus, it necessitates a novel encryption paradigm capable of maximally disentangling \usi from depression-related features, selectively obfuscating identity-sensitive attributes while preserving the integrity of depression-related features. To this end, we implement Subspace Decomposition \cite{Subspace-Decompo-1, Subspace-Decompo-2}:
\begin{equation}
S = \mathbf{U}_{\text{nd}} \alpha + \mathbf{U}_{\text{d}} \beta,
\label{eq:1}
\end{equation}

Here, \( \mathbf{U}_{\text{nd}} \in \mathbb{R}^{d \times k} \) and \( \mathbf{U}_{\text{d}} \in \mathbb{R}^{d \times m} \) are orthogonal matrices (\( \mathbf{U}_{\text{nd}}^T \mathbf{U}_{\text{d}} = 0 \)), they correspond to the orthonormal basis vectors of the NON-depressive feature subspace and the depression-specific feature subspace, respectively. \( \alpha \in \mathbb{R}^k \) and \( \beta \in \mathbb{R}^m \) are coefficient vectors.

Note that \(\mathbf{U}_{\text{nd}}\) is defined as the NON-depressive feature subspace rather than a dedicated \usi subspace, due to potential overlaps between depression-related and identity-sensitive attributes. Our method ensures that depression-specific features are retained in \(\mathbf{U}_{\text{d}}\), while enforcing orthogonality constraints between \(\mathbf{U}_{\text{d}}\) and the non-depressive feature subspace \(\mathbf{U}_{\text{nd}}\), which maximizes \usi-related feature retention. 

Subspace decomposition exhibits two critical properties: (1) Disentanglement, where two sub-features are maximally decoupled across orthogonal subspaces, \( \mathbf{U}_{\text{nd}}^T \mathbf{U}_{\text{d}} = 0 \), and (2) Reconstructability, enabling approximate signal recovery via \( S \approx \mathbf{U}_{\text{nd}} \alpha + \mathbf{U}_{\text{d}} \beta \). These properties jointly establish the theoretical foundation for deploying subspace decomposition in our framework.

\subsection{Differential Privacy} \label{DiffPri}
Differential Privacy (\ie, DP) \cite{Differential-Privacy-1, Vasa04052023} is a rigorous mathematical framework designed to protect individual data privacy in statistical analyses and machine learning. It ensures that the presence or absence of any single individual’s data in a dataset has a negligible impact on the algorithm's output, preventing adversaries from inferring sensitive information about specific participants.

Under Differential Privacy (DP), for two neighboring datasets D and D (which differ by exactly one record, \eg, \( D = D \cup {x} \)), the $(\epsilon, \delta)$-Differential Privacy requirement states that for all possible output subsets \( S \subseteq Range(A) \):

\begin{equation}
    \Pr[A(D) \in S] \leq e^{\epsilon} \cdot \Pr[A(D') \in S] + \delta,
\end{equation}

Here, \(\epsilon\) is the privacy budget which controls the strength of privacy guarantees. \(\delta\) is the failure probability, which allows a small probability of violating the privacy bound. And A is the randomized algorithm.

We proposed a model variant that implements Differential Privacy and conducted corresponding ablation studies. The implementation and results are reported in \ref{subsubsec: diffpri} and \ref{subsec:deepdive}

\subsection{Progressive Training}
Progressive Training refers to a strategy where multiple submodules of a model are trained in a specific sequence or priority, typically divided into multiple stages. Each stage focuses on optimizing parameters of a specific component, ultimately achieving overall performance improvement through combination or joint fine-tuning. Its strength lies in progressive enhancement — gradually introducing complex functionalities across stages — and dynamic adaptability, which allows flexible adjustments to subsequent training strategies based on intermediate results.

Progressive Learning is commonly employed in adversarial training, such as in Generative Adversarial Networks (\ie, GANs), where the generator and discriminator are alternately trained to compete and learn from each other \cite{GAN-1}. Additionally, it is applied to Cascaded Models, where submodels are trained sequentially following a processing pipeline, and to Mixture of Experts (\ie, MoE), where multiple submodels ("experts") specialize in different input subspaces and dynamically combine their outputs through a gating mechanism \cite{william2022switch}.

In our implementation, the model involves multiple loss functions and components serving distinct purposes. To optimize training, we adopt Progressive Learning, leveraging its capability for progressive enhancement, and dynamically adjust training configurations based on task-specific requirements to maximize performance.

%% file: chapters/chapter5-SystemDesignandMethodology.tex
\section{System Design} \label{sec:SysDes}


This section presents the proposed scheme design, which utilizes the Subspace Decomposition of two heterogeneous features. The framework remains secure and trustable even when Party B attempts to illegally access or misuse the data.  

\subsection{System Overview}
\begin{figure*}
    \centering
    \includegraphics[width=1\linewidth]{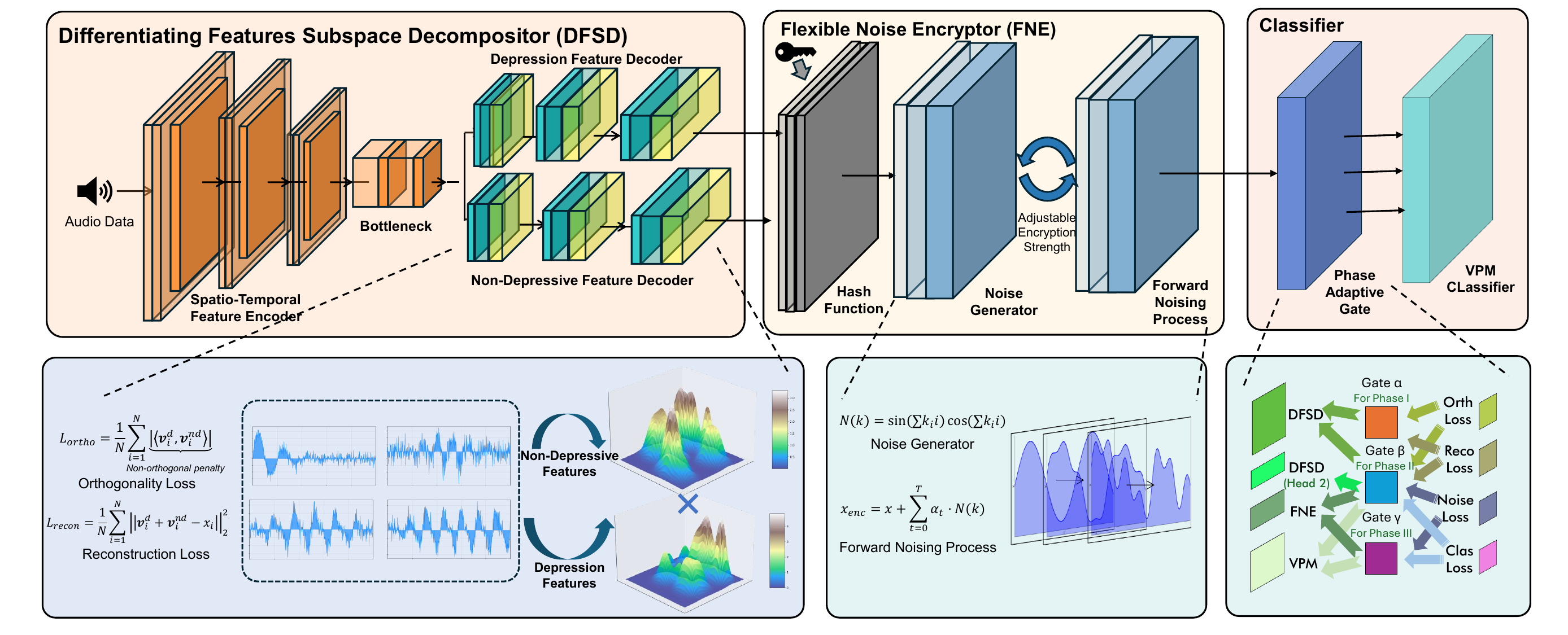}
    \caption{General design of our proposed model.}
    \label{fig:img2}
\end{figure*}

The general design of our model is shown by \ref{fig:img2}, which is composed of Three main parts: Subspace Decomposition Auto Encoder, Noise-Based Encrpytor and VPM Classifier.

\textbf{Subspace Decomposition Auto Encoder} The SDAE is designed to disentangle raw audio inputs into two mutually orthogonal latent subspaces, isolating Depression-related features and the non-depressive features by enforcing orthogonality constraints during training. It retains the ability to reconstruct the original input through feature recombination and further supports privacy-preserving data encryption by selectively perturbing non-depressive components via encryption.

\textbf{Adjustable Noise Encryptor} Drawing inspiration from Denoising Diffusion Probabilistic Models (DDPMs), our encryptor employs a controlled hierarchical noise infusion mechanism designed to fulfill three critical requirements: (1) ensuring secure data transformation, (2) maintaining model-processable signal integrity, and (3) enabling adjustable encryption intensity. Mirroring the progressive noise scheduling inherent to DDPMs, this mechanism strategically introduces structured perturbations across sequential diffusion steps. 

\textbf{VPM Classifier} For the final classification layer, we adhere to the architectural configuration proposed in the Voice-based Pretrained Model (VPM) framework: a simple fine-tuning network is directly attached to the pretrained feature extractor to serve as the ultimate classification head. This design deliberately avoids complex architectural modifications, retaining only essential projection layers for task-specific adaptation. Empirical results demonstrate that within our \name framework, such a straightforward implementation suffices to achieve the desired performance.



\label{subsec:DatPre}
During the preprocessing stage, we implement a three-phase audio preprocessing pipeline. At the foundational phase, we perform uniform amplitude scaling based on peak values:

$$
x_{norm}=\frac{x}{\max(|x|)}
$$

This operation linearly maps each audio sample to the [-1, 1] range while preserving dynamic relationships. We adopt this scaling to address inherent volume variations across datasets and between individual recordings. This scaling process can be conceptually regarded as a specialized form of normalization.

In phase II, we selectively extract participant-specific vocal segments from raw audio and perform duration-adjusted intra-subject reassembly — exclusively recombining speech clips from the same individual based on their temporal lengths, such that each recombined clip has an approximate duration of 2 second. This step filters out noises and interviewer speech artifacts, adapted from \cite{huang2024depression}, with a slight modification whereby the recombination yields clips with an approximate duration of 2 seconds. The details will be introduced in Section \ref{subsec:DatPre2}.

In Phase III, we implement temporal resampling on all recombined audio segments to achieve uniform time-series lengths across samples. This temporal standardization ensures extracted digital representations maintain identical temporal dimensions—a prerequisite for consistent batch processing in deep neural networks.

\subsection{Subsapce Decomposition Auto Encoder}
Subsapce Decomposition Auto Encoder \cite{Subspace-Decompo-3, Autoencoder-1} (Hereafter refer as SDAE) is a specialized variant of autoencoders that integrates subspace decomposition techniques. Serving as the "gateway component" of our network architecture, it operates at the highest hierarchical level to decompose input data into orthogonal subspaces, thereby enabling efficient feature disentanglement.

The theoretical foundation of the Subspace Decomposition Autoencoder (SDAE) is illustrated in Equation \ref{eq:3}. Here, we define the notation and their meanings: \( \mathbf{U}_{\text{nd}} \in \mathbb{R}^{d \times k} \) is the Non-depressive feature. \( \mathbf{U}_{\text{d}} \in \mathbb{R}^{d \times m} \) is the Depression-specific feature (note that $ \mathbf{U}_{\text{nd}} $ is not the PII feature, a distinction clarified in Section \ref{sec:PreObs}). \( \alpha \in \mathbb{R}^k \) and \( \beta \in \mathbb{R}^m \) are coefficient vectors, \( \epsilon \) is the noise term. Based on our model’s architectural design, we set \( k=m \). Consequently, we define \( \alpha=\beta=\vec{\bm{1}} \).

\begin{equation}
S = \mathbf{U}_{\text{nd}} \alpha + \mathbf{U}_{\text{d}} \beta + \epsilon,
\label{eq:3}
\end{equation}

Prior to inputting audio into the model, specialized preprocessing is required, including target speech extraction, duration-based segmentation, and reintegration to meet the model’s input specifications, as introduced in \ref{subsec:DatPre}.

Within the Spatio-Temporal Feature Encoder, the audio input is segmented and processed synchronously. This strategy reduces computational complexity, lowers memory requirements during training, and accelerates convergence. The architecture employs three consecutive fully-connected (FC) layers interleaved with nonlinear activation modules, coupled with Residual Blocks (ResBlocks), to ensure precise extraction of spatio-temporal feature representations. The encoded features are subsequently propagated and refined as latent variables within the bottleneck layer.

Two feature decoders are subsequently introduced to separately extract depression-specific features (\(\mathbf{U}_{\text{d}}\)) and non-depressive features (\(\mathbf{U}_{\text{nd}}\)). These features reside in distinct subspaces that are designed to be orthogonal and uncorrelated -- a critical property aligned with the subspace decomposition framework, \( \mathbf{U}_{\text{nd}}^T \mathbf{U}_{\text{d}} = 0 \). Still, an additional requirement — reconstructability — necessitates that the feature vectors from both subspaces can be effectively recombined to reconstruct the original audio. To jointly enforce these dual constraints, two loss functions are introduced within the optimization framework:

\begin{equation}
\text{L}_{\text{ortho}} = \frac{1}{N}\sum_{i=1}^{N}
\underbrace{\left| \langle \bm{v}_i^{(d)}, \bm{v}_i^{(nd)} \rangle \right|}_{\mathclap{Non-orthogonal\ penalty}},
\label{eq:4}
\end{equation}

\begin{equation}
\text{L}_{\text{recon}} = \frac{1}{N}\sum_{i=1}^{N}
{\|\bm{v}_i^{(d)}+\bm{v}_i^{(nd)}-x_i\|}_2^2,
\label{eq:5}
\end{equation}

In Equation \ref{eq:4}, \( \bm{v}_i^{(d)} \) and \( \bm{v}_i^{(nd)} \) represent the depression and non-depressive features of each sample in subspaces \( \mathbf{U}_{\text{d}} \) and \( \mathbf{U}_{\text{nd}} \) (\( \mathbf{U}_{\text{d}}=span\{\bm{v}_1^{(d)},\bm{v}_2^{(d)},\cdots,\bm{v}_N^{(d)}\} \), Similarly for \( \mathbf{U}_{\text{nd}} \)). This Orthogonality Loss is designed to enforce \( \mathbf{U}_{\text{d}} \) and \( \mathbf{U}_{\text{nd}} \) to form mutually orthogonal subspaces.

\begin{lemma}[Geometric Orthogonality] \label{lem:geo_ortho} \cite{spacetime-1}
Given two orthogonal matrices $ \mathbf{Q}_a, \mathbf{Q}_b \in \mathbb{R}^{n \times k} $,
the following statements are equivalent:
\begin{enumerate}
    \item $ \mathbf{Q}_a^\top \mathbf{Q}_b = \mathbf{0} $ 
    \item $ \mathbf{Q}_a \perp \mathbf{Q}_b $
    \item $ \forall \bm{u} \in \mathcal{R}(\mathbf{Q}_a), \bm{v} \in \mathcal{R}(\mathbf{Q}_b): \langle \bm{u}, \bm{v} \rangle = 0 $
    \item $ \mathcal{R}(\mathbf{Q}_a) \cap \mathcal{R}(\mathbf{Q}_b) = \{\bm{0}\} $
\end{enumerate}
\end{lemma}

We aim to enforce sample-level orthogonality:

\begin{equation}
    \langle \bm{v}_i^{(d)},\bm{v}_j^{(nd)}\rangle=0,\forall i,j\in{1,...,N},
\end{equation}

The current Orthogonality Loss implicitly implements a weakly-constrained formulation via the summation relaxation:

\begin{equation}
    \sum_{i,j=1}^{N}|\langle \bm{v}_i^{(d)},\bm{v}_j^{(nd)}\rangle| \to 0,
\end{equation}

Thereby promoting the subspace orthogonality \( \mathbf{U}_{\text{nd}}^T \mathbf{U}_{\text{d}} = 0 \) based on Lemma \ref{lem:geo_ortho}.\\

In Equation \ref{eq:5}, \( \bm{v}_i^{(d)} \) and \( \bm{v}_i^{(nd)} \) represent the depression and non-depressive features of each sample in subspaces \( \mathbf{U}_{\text{d}} \) and \( \mathbf{U}_{\text{nd}} \), while \( x_i \) represents the original signal of the audio sample. By minimizing the Mean Squared Error (MSE) reconstruction loss, we aim to ensure that the two sub-features decomposed by the SDAE can optimally recombine to reconstruct the original audio signal.

Both decoders adopt a three-layer FC architecture identical to the encoder, complete with nonlinear activation modules, ensuring consistent output feature dimensions across both pathways. These features will undergo phase-specific transformations — including encryption, fusion, or direct processing — depending upon the current training phase's operational requirements.

\label{subsubsec: diffpri}
\textbf{Differential Privacy} is a rigorous privacy framework that ensures the output of a data analysis algorithm does not reveal whether any single individual's data is included in the input dataset. The formal definition and background are introduced in Sec.~\ref{DiffPri}.

In this work, we explore the integration of Differential Privacy into our model by constructing a variant trained with DP-SGD (Differentially Private Stochastic Gradient Descent)~\cite{SP-3}. Specifically, we apply DP-SGD to train the SDAE module, with the goal of protecting latent variables from leaking user-specific information. This is achieved by injecting calibrated Gaussian noise during training to limit the influence of individual data points. The theoretical correctness of DP-SGD has been formally proven in prior work~\cite{Vasa04052023, SP-3}.

DP-SGD modifies the standard SGD optimization procedure in three main steps:
\begin{enumerate}
    \item Per-example gradient computation: Gradients are computed for each example individually, rather than for the entire batch.
    \item Gradient clipping: The $\ell_2$-norm of each per-example gradient is clipped to a predefined threshold $C$ to bound the influence of any single example.
    \item Noise addition: Gaussian noise, scaled by $C$, is added to the averaged gradient across the batch, ensuring that the update satisfies $(\varepsilon, \delta)$-differential privacy.
\end{enumerate}

We evaluate the impact of this DP-preserving variant through an ablation study in the experimental section \ref{subsec:deepdive}, analyzing how incorporating DP-SGD affects the performance of our model.

\subsection{Adjustable Noise Encryptor}
The proposed encryption paradigm establishes a bijective mapping between the original feature space \( \mathbf{U} \) and the obfuscated space \( \mathbf{\tilde{U}} \) through a Deterministic Tensor Diffusion mechanism. As depicted in \ref{fig:img3}, the core lies in the key-conditioned noise generation and parameterized diffusion scheduling.

Let \(x\in \mathcal{X}\) denote the raw feature vector and \( k \in \mathcal{K} \) the secret key. The encryption process implements an iterative nonlinear transformation:

\begin{equation}
    \tilde{x}^{(t+1)} = \tilde{x}^{(t)}+\alpha(t)\cdot\Phi(\Gamma(k)),
\end{equation}

where, \(\Gamma(k)=\sum_{j=1}^mk_i\cdot j\) generates a key-specific hash code. \(\Phi(\cdot)\) denotes the nonlinear projector: \(sin(\cdot)cos(\cdot)\). \(\alpha(t)=1-t/T\) defines the annealed noise scheduler.

The noise pattern \(\Phi(\Gamma(k))\) is key-dependent yey input-agnostic, ensuring that \(\mathbb{E}_k[\tilde{x}|x]=x+\epsilon,\ \epsilon\sim \mathcal{N}(0, \sigma^2 \mathbf{I})\). The perfect reversibility is also achieved when \(k\) is known. The decay factor \(\alpha(t)\) follows a curriculum annealing strategy, which theoretically guarantees convergence.\\

\begin{algorithm}
\caption{Deterministic noise-based encryption.}
\label{alg:det_tensor_diffusion}
\begin{algorithmic}[1]
\State \textbf{Input:} Feature $x$, Key $k$, Steps $T$
\State \textbf{Output:} Obfuscated feature $\widetilde{x}$
\State Initialize \(\widetilde{x}^{(0)} \gets x\)
\State Compute key hash: $\Gamma \gets \sum_{j=1}^{m} k_j \cdot j$ 
\State Generate base noise: $\Phi \gets \sin(\Gamma) \odot \cos(\Gamma)$
\For{$t = 1$ \textbf{to} $T$}
    \State $\alpha \gets 1 - t/T$
    \State $\widetilde{x}^{(t)} \gets \widetilde{x}^{(t-1)} + \alpha \cdot \Phi$
\EndFor
\State \Return $\widetilde{x}^{(T)}$
\end{algorithmic}
\end{algorithm}

\hspace{0.5em}

During the encryption process, we modulate the encryption strength by adjusting the value of parameter \(T\), thereby achieving an optimal trade-off between confidentiality and accuracy, demonstrated by subsequent experiments in section \ref{sec:ExpEva}. 

\subsection{VPM Classifier}
In terms of the output architecture, our model adopts the strategic framework from the Voice-based Pre-training Model (VPM) \cite{huang2024depression}, where a compact fine-tuning network serves as the final classification layers to generate depression prediction scores. 

Prior to the VPM Classifier, we implement a Phase-Adaptive Gating Mechanism that dynamically regulates data throughput according to three predefined training phases, capable of operating in three distinct modes: full transmission, partial transmission, and complete blockage of data flow, governed by phase-dependent policies.

In the initial two tiers of the VPM architecture, we architect a convolutional neural network (CNN) \cite{SP-4} backbone incorporating batch normalization, nonlinear activation layers, and dropout regularization. The terminal layer employs a fully-connected projection head to map the refined representations into depression classification logits.

For the classification loss, we employ the Cross Entropy loss with label smoothing, a mathematically principled choice for optimizing probabilistic predictions in depression detection:

\begin{equation}
L_C=\frac{\sum_{i=1}^N(z_{i,y_i}-log(\sum_{k=1}^Ke^{z_{ik}}))}{N},
\end{equation}

where \( z_{ik}\) represents the unnormalized logit value of the i-th sample for class \( k \), and \(y_i\) denotes the ground-truth class label. This loss function ensures stable optimization.

\subsection{Various Training Phases}
The framework is executed through three sequential training phases. In Phase I, we focus on training the Differentiating Features Subspace Decompositor (\sdae) to decompose raw audio samples into two orthogonal subspaces—depressive and non-depressive features—while strictly enforcing orthogonality and reconstruction constraints. The encryptor remains untrained as a deterministic module, where its output is fully determined by predefined encryption keys and strength parameters. During Phase II, the VPM Classifier is trained on unencrypted data, which is reconstructed directly from the two subspaces, establishing baseline diagnostic performance for raw audio processing. Finally, Phase III adapts the classifier to encrypted inputs: non-depressive features are first encrypted and then recombined with depressive features to synthesize privacy-preserving audio samples. The VPM Classifier undergoes fine-tuning via backpropagation on these encrypted samples.

\textbf{Phase I} In this training phase, we utilize three loss functions: Orthogonality Loss, Reconstruction Loss, and Classification Loss. The Orthogonality Loss and Reconstruction Loss ensure the \sdae fulfills its core functionality, while the Classification Loss enforces strong relevance between depressive features and depression detection. The combined effect of Orthogonality Loss and Classification Loss minimizes the diagnostic influence of features from the other decoder.

During this phase, the \sdae and VPM Classifier are employed. The depression-related features are fed directly into the VPM Classifier to compute the Classification Loss. While the Classification Loss is backpropagated through the entire network, the Orthogonality Loss and Reconstruction Loss gradients only update the \sdae. To prioritize \sdae training, we introduce a hyperparameter to amplify the emphasis on Orthogonality/Reconstruction Losses and attenuate the Classification Loss influence.

\textbf{Phase II} Subsequently, we proceed to Phase II. During this and subsequent phases, \sdae's parameters are frozen to preserve its orthogonality and reconstruction constraints. The depression and non-depressive features extracted by \sdae are reintegrated and fed into the VPM Classifier (without encryption). This training phase is dedicated to training a model capable of processing unencrypted data.

\textbf{Phase III} In the final Phase III, the \sdae's parameters remain frozen. The encrypted non-depressive features are fused with depression features and fed into the VPM Classifier for training, establishing a model optimized for processing encrypted clinical audio data.

%% file: chapters/chapter6-ExperimentalEvaluation.tex
\section{Experimental Evaluation} \label{sec:ExpEva}
\subsection{Dataset} \label{subsec:Dat}
We employ the dataset following established research conventions — DAIC-WOZ (Distress Analysis Interview Corpus-Wizard of Oz).

The Distress Analysis Interview Corpus (DAIC) contains clinical interviews designed to support the diagnosis of psychological distress conditions such as anxiety, depression, and post traumatic stress disorder. The interviews are conducted by humans, human controlled agents and autonomous agents, and the participants include both distressed and non-distressed individuals \cite{gratch2014DAIC}. 

\subsection{Implementation \& Hyperparameters} \label{subsec:ImpHyp}
Across different experimental configurations, certain implementation-specific hyperparameters may vary. Here we present the standardized experimental protocols and hyperparameters consistently adopted across all trials.

All experiments were conducted on a single NVIDIA GeForce RTX 4090 D GPU with 24GB VRAM, running under Python 3.9.13.

The SDAE's three-layer encoder-decoder architecture exhibits feature dimensionalities of 32,000, 8,000, and 2,000 across successive layers. The weighting factors for orthogonality loss and reconstruction loss are both set to 10, with configurations: dropout rate = 0.2, learning rate = 1e-4, and weight decay = 0.01.

The final depression classification task is formulated as a binary classification problem, where the output 0/1 indicates whether the individual doesn't have depression (0 indicates depression, 1 indicates no depression).  We set the batch size to 32, with the total number of epochs set to 10. After conducting grid experiments, we determined that the optimal threshold for evaluating the model output is 0.4 — that is, if the predicted score is greater than 0.4, it is classified as depressed; otherwise, as non-depressed. Interestingly, this threshold coincides with the proportion corresponding to the depression cut-off in the PHQ-8 score (the PHQ-8 ranges from 0 to 24, and scores above 9 are considered indicative of depression, so \(\frac{10}{25} = 0.4\)).

In contrast, the PII identification task for each data point is formulated as a pairwise comparison task, where the model is presented with two data samples and outputs whether the two audio recordings come from the same individual. This setup simulates the real-world scenario in which malicious actors exploit audio data by comparing it against a known database to determine the source of the audio.

\subsection{Data Preprocessing} \label{subsec:DatPre2}
To meet our specific experimental requirements and enhance the data, we performed preprocessing on the dataset, consisting of three stages: amplification, extraction/recombination, and resampling. In the second stage, we decomposed the audio data in DAIC-WOZ into 32,400 segments and recombined them into 9,753 audio clips. After the third stage, every clip can be loaded into digital form of length 32000.

Additionally, in the Depression Detection task, we randomly partitioned the dataset into ten different train-test splits, evaluated the model on each split, and averaged the results to ensure the robustness of the model. In the PII Identification task, we randomly paired the data, resulting in 189 positive pairs and 611 negative pairs, for a total of 800 pairs of samples.

\subsection{Metrics}
In this paper, we leverage the following metrics to evaluate the performance of the proposed architecture on depression detection:

\begin{itemize}

\item [1)] \emph{Recall.} Given a series of input data and the corresponding outcomes, the \emph{recall} can be defined as:
\begin{equation}
Recall = \frac{TP}{TP+FN},
\end{equation}
where $TP$, $FN$ represent True Positive and False Negative, respectively. Recall measures the proportion of actual positive cases that are correctly identified.

\item [2)] \emph{Precision.} Given a series of input data and the corresponding outcomes, the \emph{precision} can be expressed as:
\begin{equation}
Precision = \frac{TP}{TP+FP},
\end{equation}
where $FP$ represents the False Positive. Precision measures the proportion of predicted positive cases that are truly positive.

\item [3)] \emph{F1-score.} The \emph{F1-score} can be represented as:
\begin{equation}
\text{\emph{F1-score}} = \frac{2\times Precision\times Recall}{Precision+Recall}.
\end{equation}
The F1-score provides a harmonic mean of precision and recall, balancing the two metrics.

\item [4)] \emph{Accuracy.} Given a series of input data and the corresponding outcomes, the \emph{accuracy} can be defined as:
\begin{equation}
Accuracy = \frac{TP + TN}{TP + TN + FP + FN},
\end{equation}
where $TN$ represents True Negative. Accuracy measures the overall proportion of correctly classified cases.

\end{itemize}

In the PII Identification task, we will focus more on the other three metrics.

\begin{itemize}

\item [5)] \emph{False Acceptance Rate (FAR).} The \emph{False Acceptance Rate} is a metric used to evaluate the rate at which negative samples (i.e., non-matching cases) are incorrectly accepted as positive in a classification or authentication system. It can be defined as:
\begin{equation}
FAR = \frac{FP}{FP + TN},
\end{equation}
where $FP$ represents the False Positive, and $TN$ represents the True Negative. FAR indicates how frequently the system incorrectly identifies a negative sample as positive, which is a critical metric for security-sensitive applications like biometric authentication.

\item [6)] \emph{False Rejection Rate (FRR).} The \emph{False Rejection Rate} measures the rate at which positive samples (i.e., matching cases) are incorrectly rejected by the system. It can be expressed as:
\begin{equation}
FRR = \frac{FN}{FN + TP},
\end{equation}
where $FN$ represents the False Negative, and $TP$ represents the True Positive. FRR indicates how often the system fails to correctly identify positive samples, which is crucial for user experience in systems such as facial recognition or fingerprint scanning.

\item [7)] \emph{Equal Error Rate (EER).} The \emph{Equal Error Rate} is a metric commonly used in biometric systems, such as speaker recognition or face recognition. EER is the point at which the False Acceptance Rate (FAR) equals the False Rejection Rate (FRR). It represents a threshold where both types of errors occur at the same rate. Mathematically, it can be described as:
\begin{equation}
EER = \text{FAR} = \text{FRR},
\end{equation}
where FAR is the rate of incorrectly accepting a negative sample as positive, and FRR is the rate of rejecting a true positive. The EER provides a single value to evaluate the trade-off between false positives and false negatives at the operating point where they are balanced.

\end{itemize}

These three metrics are used because, in the speaker recognition task, we focus on balancing the trade-off between false acceptances (FAR) and false rejections (FRR), which directly impact both security and user experience. FAR and FRR are more relevant than Precision and Recall, as they better reflect the system's ability to correctly handle authorized and unauthorized users. At the same time, we will also report the Accuracy result.

\subsection{Evaluation of Depression Detection}

The evaluations in this part primarily assess the performance of \name on depression detection task, which contain depression detection on unencrypted data and on encrypted data. Since the effect of encryption predominantly depends on encryption strength and keys, we evaluate our modal on 3 different encryption strengthes from each of the 4 large-size keys across the 12 experimental scenarios. To ensure the full representation of experimental effects in unfamiliar environments, the data used for each testing session are not included in the training set.

To assess the usability of \name, we first evaluate the model on unencrypted audios. As stated in \ref{subsec:ImpHyp}, for each evaluation, we set the batch size to 32, the total epochs to 10, and the threshold to 0.4. The average results of cross validation are reported in Tab. \ref{tab:EvaRes1}. The experimental findings demonstrate that the proposed arthitecture effectively detects depression with unencrypted audios.

The subsequent part of the evaluation focuses on the performance of the model on encrypted audios. We evaluate our model under three encryption strengths: 5, 10, and 25, and report the average results obtained over experiments conducted with four different keys for each strength.  The average precision is $86.5\%$, and the average recall is $86.2\%$. The experiments confirm the system's capability to accurately identify large-size keys from a series of acoustic inputs. The recognition of the Space and Enter keys exhibits slightly higher accuracy compared to other keys, attributed to their distinctive timbre. Although occasional identification errors may occur, these samples can be easily deleted due to their significant offset in location estimation.

Next, we assess the performance of our model against other methods as part of an ablation study to validate the effectiveness of our approach. Specifically, we conduct experiments using two representative techniques: Chaos Maps-Based Encryption \cite{albahrani2021review, SP-5} and Homomorphic Encryption \cite{nguyen2024efficient}. Following the same experimental setup as before, we report their Accuracy, Recall, Precision, and F1 scores for comparison in Tab. \ref{tab:EvaRes1}.

\begin{table}[t]\small
\begin{threeparttable}
\renewcommand\arraystretch{1.2}
    \centering
    \caption{Evaluation results on depression detection task.}
    \begin{tabular}{ccccc}
    \toprule[1.5pt]
        Our Arthitecture - \name & ACC & REC & PRE & F1\\ \midrule[1.5pt]
        Unencrypted & \textbf{84.75\%} & 0.88 & \textbf{0.86} & 0.87 \\ \midrule
        Encrypted (Strength 5) & 80.17\% & 0.88 & 0.85 & 0.86 \\ \midrule
        Encrypted (Strength 10) & 83.62\% & \textbf{0.92} & 0.84 &\textbf{ 0.88} \\ \midrule
        Encrypted (Strength 25) & 78.28\% & 0.91 & 0.8 & 0.85 \\ \midrule[1.5pt]
        Other Methods & ACC & REC & PRE & F1\\ \midrule[1.5pt]
        Chaos Maps-Based Encryption & 64.00\% & 0.72 & 0.83 & 0.77 \\ \midrule
        Homomorphic Encryption & 84.23\% & 0.88 & 0.86 & 0.87 \\ \bottomrule
    \end{tabular}
    \label{tab:EvaRes1}
    \vspace{0.15in}
\end{threeparttable}
\end{table}

\begin{figure}
    \centering
    \includegraphics[width=0.9\linewidth]{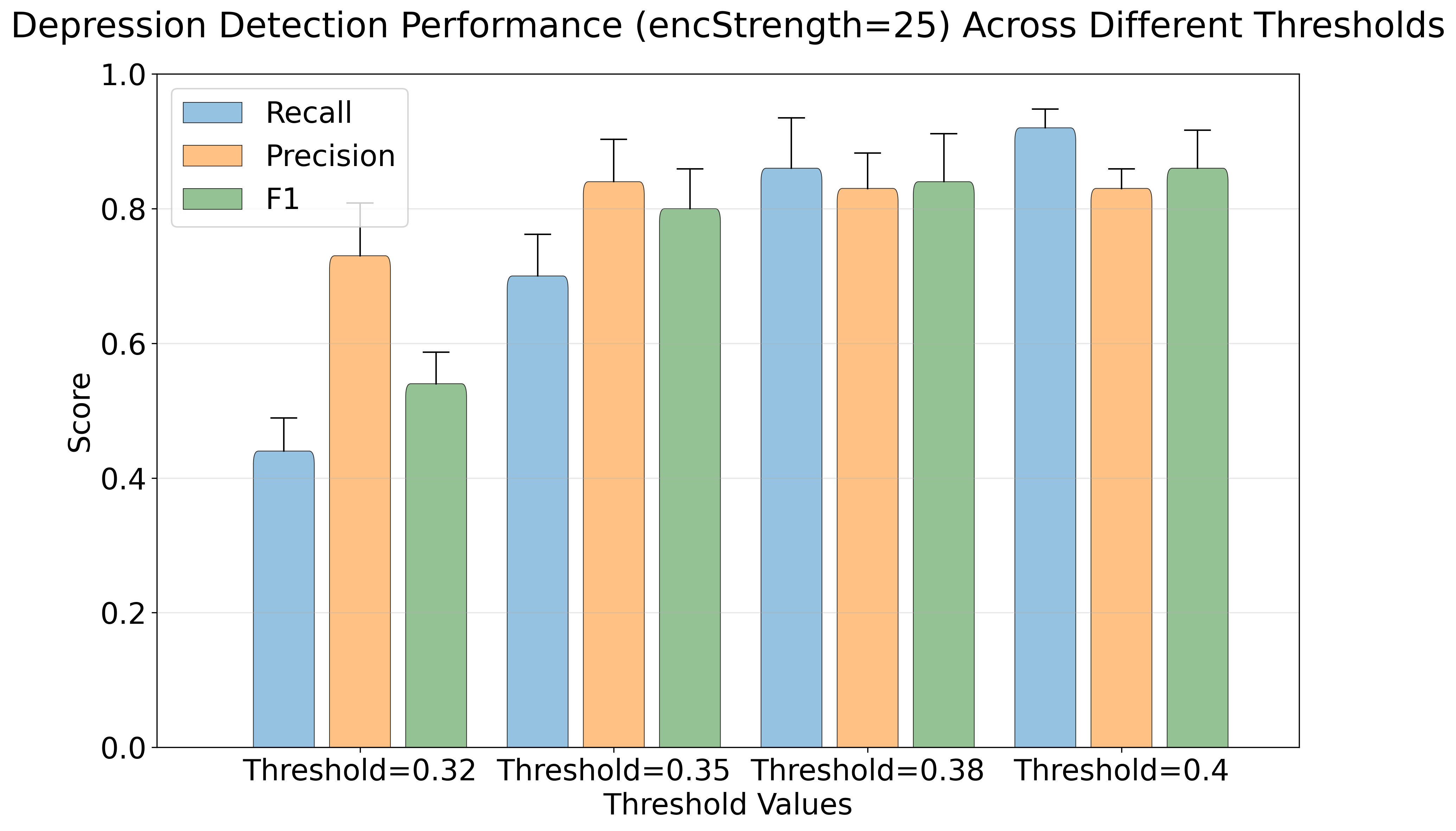}
    \caption{Confusion matrix on depression detection with encStrength 25.}
    \label{fig:img5}
\end{figure}

\begin{figure}
    \centering
    \includegraphics[width=0.9\linewidth]{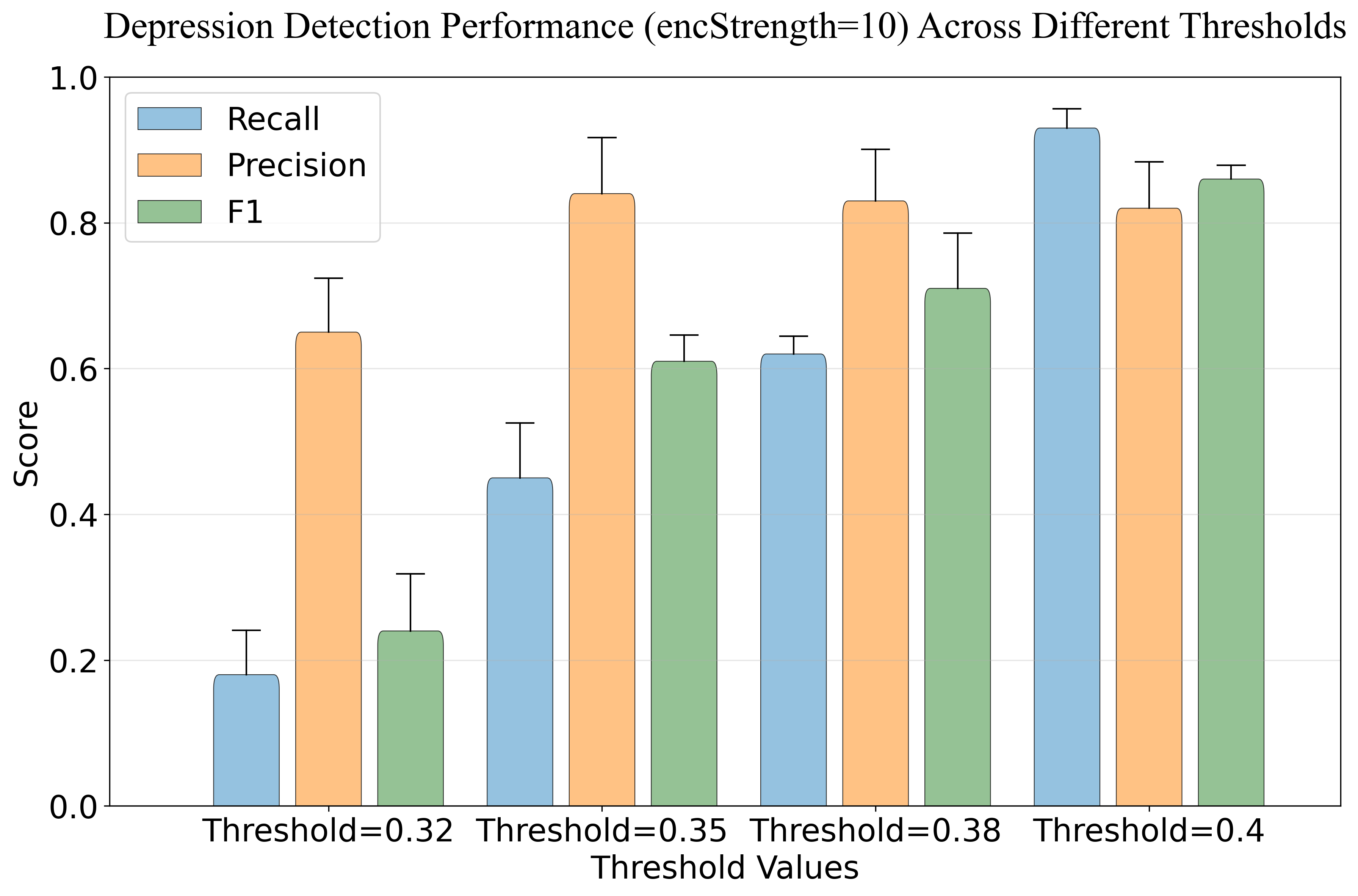}
    \caption{Confusion matrix on depression detection with encStrength 10.}
    \label{fig:img7}
\end{figure}

Finally, we also report the confusion matrices under four different thresholds at encryption strengths of 10 and 25, as shown in Fig. \ref{fig:img5} and Fig. \ref{fig:img7}, to further analyze the model’s classification behavior and demonstrate how its performance varies with different threshold settings.

As shown in Table \ref{tab:EvaRes1}, our model achieves an accuracy of 84.75\% and an F1 score of 0.87 on the original (unencrypted) data, which represents a strong performance. After applying our Encryptor structure, the model maintains comparable performance under a weak Encryption Strength of 5, achieving an accuracy of 80.17\%, with an F1 score nearly equivalent to that on the unencrypted data. Even under a strong Encryption Strength of 25, the model still achieves 78.28\% accuracy and an F1 score of 0.85. These results demonstrate that our model meets the Accuracy requirement outlined in the Three Major Challenges. Moreover, users can dynamically adjust the encryption strength to balance between model accuracy and data confidentiality. 

It is worth noting that as the encryption strength increases, the remaining depression-related information in the PII subspace becomes increasingly obfuscated. As a consequence, we observe that the model tends to predict samples as non-depressed. Combined with the issue of class imbalance in the dataset—where non-depressive samples greatly outnumber depressive ones—this bias leads to an increase in recall but a drop in precision, with recall consistently remaining high.

As shown in Table \ref{tab:EvaRes1}, when applying Chaos Maps-Based Encryption, the accuracy drops to only 64\%. This demonstrates that merely using existing encryption methods leads to complete obfuscation of feature information, making it impossible for the model to learn meaningful patterns from the data. While Homomorphic Encryption achieves strong performance on the Depression Detection task, the comparison in Fig. \ref{fig:img6} reveals that it fails to prevent malicious actors from stealing PII.


\subsection{Performance of PII Extraction}

We first report the statistical differences between the encrypted audio, the reconstructed audio, and the original audio in Tab. \ref{tab:EncRecOri}. The reported metrics include MSE, MAE, and PSNR. A lower MSE/MAE indicates higher similarity. PSNR (Peak Signal-to-Noise Ratio) represents the ratio of signal power to noise power. A higher PSNR indicates less distortion and better quality. 

From Tab. \ref{tab:EncRecOri}, it can be observed that audio encrypted using the Chaos Maps-based method differs the most from the original audio, offering the highest level of confidentiality. In contrast, audio encrypted using the Homomorphic method shows less deviation from the source data but achieves the best reconstruction fidelity. Our method lies between these two extremes.

\begin{table}[t]\small
\renewcommand\arraystretch{1.2}
    \centering
    \caption{Statistical differences between the encrpyted audio, the reconstructed audio and the original audio.}
    \begin{tabular}{ccccc}
    \toprule[1.5pt]
        Data - Method & MSE & MAE & PSNR \\ \midrule[1.5pt]
        Encrypted (\name) & 7.56 & 2.75 & -30.02dB \\ \midrule
        Reconstructed (\name) & 2.7e-4 & 1.0e-2 & 14.46dB \\ \midrule[1.5pt]
        Encrypted (Chaos Maps-Based) & 1.4e+7 & 2.7e+3 & -1.91dB \\ \midrule
        Reconstructed (Chaos Maps-Based) & 82.90 & 0.55 & 50.3dB \\ \midrule[1.5pt]
        Encrypted (Homomorphic) & 0.91 & 0.95 & -20.82dB \\ \midrule
        Reconstructed (Homomorphic) & 0.00 & 0.00 & 98.77dB \\ \bottomrule
    \end{tabular}
    \label{tab:EncRecOri}
    \vspace{0.15in}
\end{table}

Next, we evaluate the model’s performance on the PII Extraction task following the experimental setup described in Section \ref{subsec:ImpHyp}. with a slight modification: multiple audio samples from the same individual are concatenated into a longer audio segment (ranging from 30 seconds to 1 minute), as the model used for this task requires longer inputs to achieve reliable comparison accuracy. The model used is the comparison model provided by SpeechBrain \cite{ravanelli2021speechbrain}. The dataset is divided into 378 pairs, among which 189 pairs contain two audio samples from the same individual, while the remaining 189 pairs are composed of samples from different individuals. We report the model’s performance in terms of Accuracy, FAR, FRR, and EER in Tab. \ref{tab:EvaRes2}, with the threshold set to 0.85.

\begin{table}[t]\small
\renewcommand\arraystretch{1.2}
    \centering
    \caption{Evaluation results on PII extraction task.}
    \begin{tabular}{ccccc}
    \toprule[1.5pt]
        Our Arthitecture - \name & ACC & FAR & FRR & EER\\ \midrule[1.5pt]
        Encrypted (Strength 5) & 75.40\% & 0.48 & 0.01 & 0.26 \\ \midrule
        Encrypted (Strength 10) & 58.47\% & 0.83 & 0.01 & 0.3 \\ \midrule
        Encrypted (Strength 25) & \textbf{50.53\%} & \textbf{0.97} & 0.02 & \textbf{0.49} \\ \midrule[1.5pt]
        Other Methods & ACC & FAR & FRR & EER\\ \midrule[1.5pt]
        Unencrypted & 100.00\% & 0 & 0 & 0 \\ \midrule
        Chaos Maps-Based Encryption & 51.06\% & 0.75 & \textbf{0.23} & 0.35 \\ \midrule
        Homomorphic Encryption & 99.74\% & 0.01 & 0 & 0.01 \\ \bottomrule[1.5pt]
    \end{tabular}
    \label{tab:EvaRes2}
    \vspace{0.15in}
\end{table}

We first use the trained SDAE and the Encryptor under different encryption strengths (5, 10, and 25) to encrypt the data. The encrypted data is then passed to the verification model for prediction. The results are shown in Table III. It is worth noting that under full encryption, the verification model tends to classify all audio samples as belonging to the same individual, as the signals become completely randomized and chaotic — which causes the accuracy to fluctuate around 50\% at worst.

Next, we feed the original, unencrypted audio into the model. This step is intended to demonstrate that the verification model possesses strong discriminative power and can accurately determine whether two audio samples belong to the same individual when the data is not obfuscated.

\begin{figure}
    \centering
    \includegraphics[width=1\linewidth]{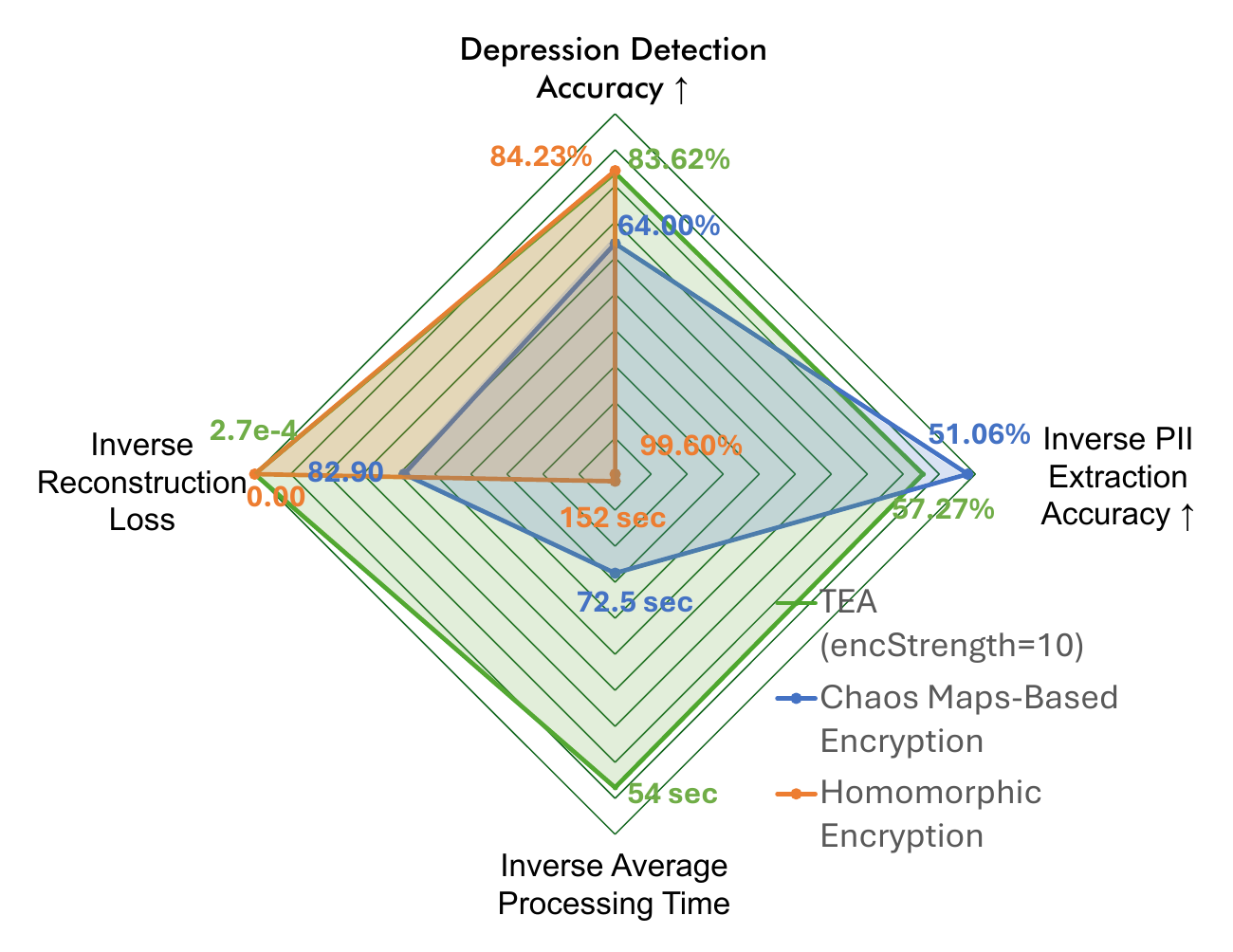}
    \caption{Radar chart comparison of the three encryption methods.}
    \label{fig:img6}
    \caption*{\footnotesize Since lower values of PII Extraction Accuracy, Reconstruction Loss, and Average Processing Time indicate better performance, we present their \textbf{inverse} values in the figure to make the comparison more intuitive.}
\end{figure}

Finally, we also present the performance of the two aforementioned encryption methods on the PII Extraction task within Tab. \ref{tab:EvaRes2}, to compare it with their performance on the Depression Detection task. In addition, we provide a visual comparison in Fig. \ref{fig:img6} to offer a more intuitive understanding of the Pros and Cons.

One of the primary contributions of our work is the ability to ensure the obfuscation and invisibility of PII within the data, and the results in Tab. \ref{tab:EvaRes2} demonstrate that our model achieves this goal. When the data is unencrypted, the model achieves 100\% accuracy and an EER of 0, indicating that the SpeechBrain model can perfectly identify whether two audio samples come from the same individual in a clean setting. However, after encryption, we observe a decrease in accuracy and a significant increase in EER as the encryption strength increases. At encryption strength of 25, the accuracy reaches 50.53\%, and the EER increases to 0.49. A detailed analysis reveals that after encryption, the model classifies nearly all audio pairs as coming from the same individual, due to the encrypted noises being highly similar to each other. As a result, the FRR remains low, while the FAR becomes extremely high, and the accuracy fluctuates around 50\% at worst.

As shown in the Tab. \ref{tab:EvaRes2} and Fig. \ref{fig:img6}, the results obtained from data encrypted using Chaos Map-Based Encryption are comparable to those achieved by our model under high encryption strength. This demonstrates that our model provides a level of PII protection similar to traditional encryption methods. However, unlike those methods, our model still retains the ability to extract depression-related features from the encrypted audio. In contrast, the accuracy and EER after Homomorphic Encryption remain close to those of the unencrypted case, indicating that although Homomorphic Encryption achieves good performance on the Depression Detection task, it fails to adequately protect PII information.

\subsection{Deep Dive into \name}\label{subsec:deepdive}
The main purpose of this section is to demonstrate the impacts of various factors, e.g., environmental factors, dataset-related factors, and data processing factors, on the \name system. To isolate the impact of individual factors on the experimental results and effectively control variables, the experiments in this section are conducted with the encryption strength fixed at 10 and the threshold fixed at 0.4. Since the following experiments are not strongly related to the main achievements of our model, they are conducted using only a single key on three splits.

\subsubsection{Impact of DP-SGD}
In this experimental setup, we conducted experiments on both the DP and non-DP versions using the configurations described above. During the experiments, we observed that the DP version had significantly longer processing times; therefore, we additionally report the average training time of the models in the table. The experimental results are reported in Tab. \ref{tab:EvaResDP}.

\begin{table}[t]\small
\centering
\begin{threeparttable}
\renewcommand\arraystretch{1.2}
    \caption{Evaluation results on DP vs. non-DP.}
    \begin{tabular}{lccccc}
    \toprule[1.5pt]
          & ACC & REC & PRE & F1 & Time \dag \\ \midrule[1.5pt]
        non-DP & 81.40\% & 0.96 & 0.84 & 0.90 & 24min \\ \midrule
        DP & 76.00\% & 0.87 & 0.83 & 0.85 & 112min \\ \bottomrule
    \end{tabular}
    \begin{tablenotes} 
    \item[\dag] Here we report the average time required to train for 10 epochs across different splits.
    \end{tablenotes}
    \label{tab:EvaResDP}
    \vspace{0.15in}
\end{threeparttable}
\end{table}

From the table, it can be seen that after applying DP, the average processing time increased by 4 to 5 times, while the accuracy decreased but remained within an acceptable range. Since DP enhances security under the differentially private setting, it can be offered as an optional feature for users.


\subsubsection{Impact of Noises}

In the DAIC-WOZ dataset, each original audio contains the participant's speech, the interviewer's speech, and background noise. Through data preprocessing, we removed the interviewer's voice and background noise from the audio, ensuring a strong correlation between the data and the participant in our experiments.

To evaluate the system's robustness against noise, we conduct experiments under three different scenarios with varying noise levels. The scenarios and corresponding noise types are detailed in Tab. \ref{tab:SceTyp}. Specifically, the noise intensity ranges from $43.31dB$ to $46.36dB$. The results are reported in Tab. \ref{tab:EvaResNoise}

\begin{table*}[t]
\renewcommand\arraystretch{1.2}
    \centering
    \caption{Scenarios and types of noises.}
    \resizebox{0.8\textwidth}{!}{
    \begin{tabular}{cc}
    \toprule[1.5pt]
        Scenarios & Type of noises \\ \midrule[1.5pt]
        Clean & None \\ \midrule
        Environmental Noise & Ambient environmental noise (46.36dB, 10\%) \\ \midrule
        Speech Noise & Ambient environmental noise (46.36dB, 10\%), Interviewer's speech noise (43.31dB, 10\%) \\ \bottomrule
    \end{tabular}}
    \label{tab:SceTyp}
    \vspace{0.15in}
\end{table*}

\begin{table}[t]\small
\centering
\begin{threeparttable}
\renewcommand\arraystretch{1.2}
    \caption{Evaluation results on noises}
    \begin{tabular}{lcccc}
    \toprule[1.5pt]
          & ACC & REC & PRE & F1 \\ \midrule[1.5pt]
        Clean & 81.4\% & 0.96 & 0.84 & 0.90 \\ \midrule
        Environmental Noise & 79.88\% & 0.96 & 0.8 & 0.87 \\ \midrule
        Speech Noise & 75.2\% & 0.99 & 0.68 & 0.81 \\ \bottomrule
    \end{tabular}
    \label{tab:EvaResNoise}
    \vspace{0.15in}
\end{threeparttable}
\end{table}

As shown in the table, the accuracy slightly decreases after introducing environmental When more third-party noise is introduced, the accuracy drops substantially. Through analysis, we find that the model tends to classify the interviewer’s voice as non-depressive, leading to almost all samples being predicted as non-depressive.

\subsubsection{Impact of Various Audio Lengths}
In the original experimental setup, each audio was recombined to approximately 2 seconds. In this experiment, we evaluate the impact on accuracy when the audio is recombined to different lengths. The results are reported in Tab. \ref{tab:EvaResLength}

\begin{table}[t]\small
\centering
\begin{threeparttable}
\renewcommand\arraystretch{1.2}
    \caption{Evaluation results on various audio lengths}
    \begin{tabular}{lcccc}
    \toprule[1.5pt]
          & ACC & REC & PRE & F1 \\ \midrule[1.5pt]
        2 sec & 81.4\% & 0.96 & 0.84 & 0.90 \\ \midrule
        3 sec & 80.6\% & 0.96 & 0.83 & 0.89 \\ \midrule
        5 sec & 81.2\% & 0.96 & 0.84 & 0.90 \\ \bottomrule
    \end{tabular}
    \label{tab:EvaResLength}
    \vspace{0.15in}
\end{threeparttable}
\end{table}

\begin{figure}
    \centering
    \includegraphics[width=0.8\linewidth]{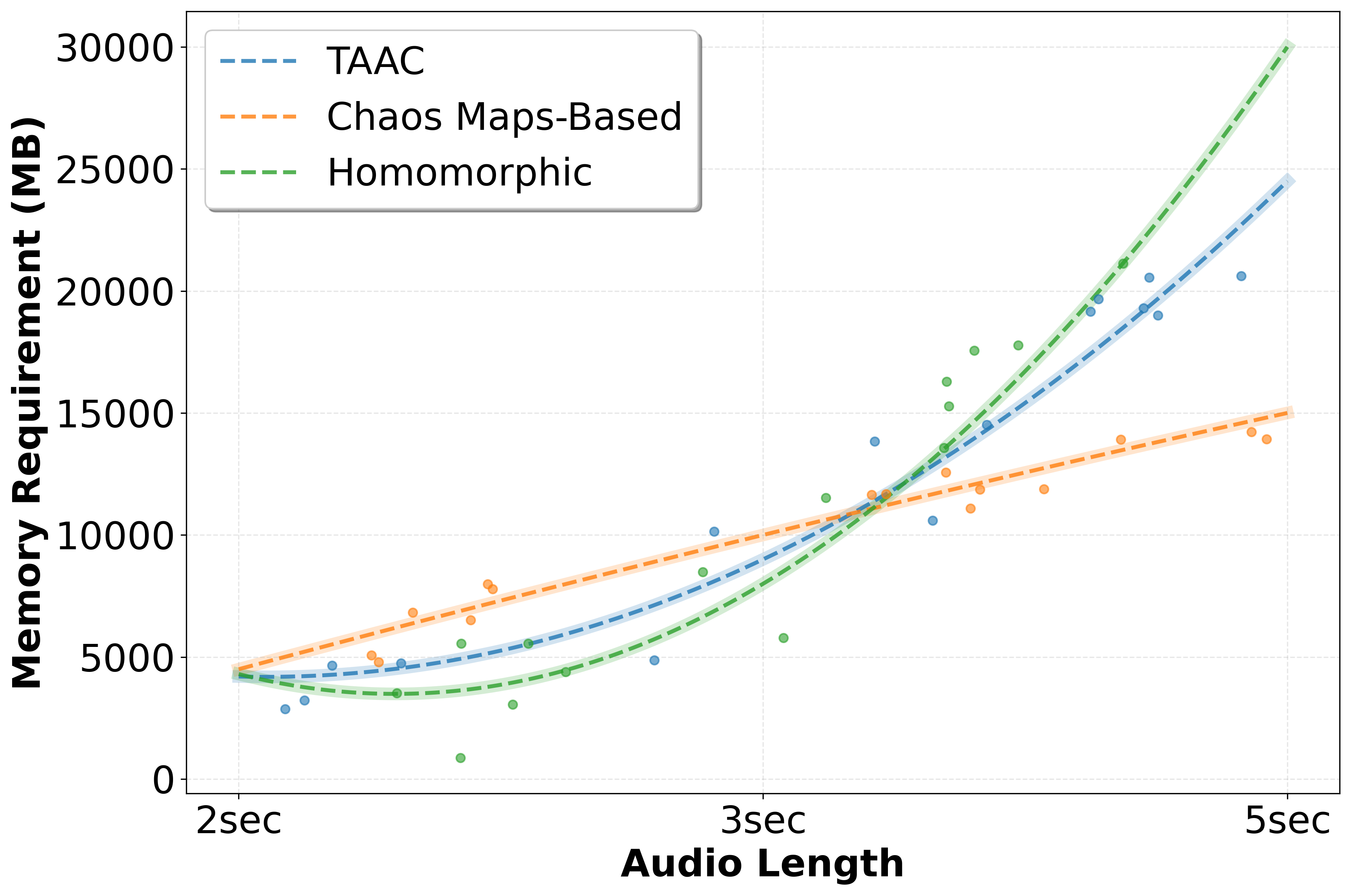}
    \caption{Changes of GPU memory requirements under different audio lengths (unit: MB).}
    \label{fig:GPULen}
\end{figure}

As shown in the Tab. \ref{tab:EvaResLength}, the experimental results are barely affected by different audio lengths, indicating that under relatively short durations, varying audio lengths do not have a significant impact on our model. However, it is worth noting that as the audio length increases, the training cost grows exponentially, and our training hardware can only support processing up to a maximum audio length of 5 seconds. We present in Fig. \ref{fig:GPULen} the changes in GPU memory requirements under different audio lengths.

%% file: chapters/chapter7-Conclusion.tex
\section{Conclusion} \label{sec:Con}

This paper presents a practical and trustworthy data processing framework, \name, for performing depression detection on audio data within a trustable setting. To address the trade-off between encryption and accuracy, we leverage Subspace Decomposition and propose a novel architecture, SDAE. Extensive experimental results demonstrate that, compared with existing purely encryption-based methods such as Chaos Maps-Based Encryption and Homomorphic Encryption, the proposed system effectively balances accuracy and security—whereas prior methods tend to ensure only one. Furthermore, by drawing inspiration from diffusion models, we introduce a Adjustable Noise Encryptor, enabling \name to preserve the original data structure post-encryption. This allows the encrypted audio to remain processable by the model. Additionally, the framework incorporates an adjustable Encryption Strength parameter, granting users the flexibility to balance accuracy and privacy, thereby enhancing the system's practicality. By varying key experimental parameters, we further validate how different Encryption Strength and Threshold settings impact both the Depression Detection and PII Extraction tasks. With the integration of a VPM Classifier and the Progressive Training method, the classification performance is further improved. Even when handling noisy audio data or samples of varying lengths and sources, our model demonstrates robust and reliable performance.

%% file: chapters/Chapter-9-EthicConsideration.tex

Our work discusses the risk that user audio data in depression-detection systems may be misused by service providers. This concern is not a novel disclosure; rather, it reflects a widely recognized issue that has attracted substantial attention from stakeholders, with existing national policies already restricting abusive data practices.

The primary contribution of our paper is to introduce a technical approach that fundamentally eliminates the possibility of such misuse. Following publication, we welcome engagement with developers and vendors and encourage the adoption of our approach in real-world systems. Through this work, we intend to further emphasize to the public the importance of safeguarding audio data and to encourage vendors to critically re-examine their own security practices.


Our study makes use of the DAIC-WOZ dataset, a publicly available corpus of clinical interviews designed to support the diagnosis of psychological distress. The dataset was originally collected by the University of Southern California (USC) Institute for Creative Technologies. According to the original publications and our investigation, the data collection procedure received approval from the corresponding Institutional Review Board (IRB), and all participants provided informed consent.

Although the dataset is publicly accessible, we adhered strictly to ethical guidelines for handling sensitive data. We clearly communicated our research intentions to the dataset provider and obtained access only after approval of our application. In our analysis, we used solely anonymized audio data and relied on anonymized participant identifiers. No data linking participants to their real-world identities was exposed or processed at any stage of the study.

%% file: bib/KeyStroke.bib
@misc{ravanelli2021speechbrain,
  title     = {SpeechBrain: A General-Purpose Speech Toolkit},
  author    = {Ravanelli, Mirco and Parcollet, Titouan and Plantinga, Peter and Rouhe, Aku and Cornell, Samuele and Lugosch, Loren and Subakan, Cem and Dawalatabad, Nauman and Heba, Abdelwahab and Zhong, Jianyuan and Chou, Ju-Chieh and Yeh, Sung-Lin and Fu, Szu-Wei and Liao, Chien-Feng and Rastorgueva, Elena and Grondin, François and Aris, William and Na, Hwidong and Gao, Yan and others},
  year      = {2021},
  archiveprefix = {arXiv},
  eprint    = {2106.04624},
  primaryclass = {eess.AS},
  url       = {https://arxiv.org/abs/2106.04624},
}

@article{ye2021multi,
  author  = {Ye, Jiayu and Yu, Yanhong and Wang, Qingxiang and Li, Wentao and Liang, Hu and Zheng, Yunshao and Fu, Gang},
  title   = {Multi-modal depression detection based on emotional audio and evaluation text},
  journal = {Journal of Affective Disorders},
  year    = {2021},
  volume  = {295},
  pages   = {904--913},
  doi     = {10.1016/j.jad.2021.08.090},
  url     = {https://www.sciencedirect.com/science/article/pii/S0165032721008958}
}

@inproceedings{shen2022automatic,
  author    = {Shen, Ying and Yang, Huiyu and Lin, Lin},
  title     = {Automatic Depression Detection: An Emotional Audio-Textual Corpus and a GRU/BiLSTM-Based Model},
  booktitle = {Proceedings of the 2022 IEEE International Conference on Acoustics, Speech and Signal Processing (ICASSP)},
  year      = {2022},
  pages     = {6247--6251},
  publisher = {IEEE},
  doi       = {10.1109/ICASSP43922.2022.9746569}
}

@inproceedings{ringeval2019AVEC,
  author = {Ringeval, F. and Schuller, B. and Valstar, M. and Cummins, N. and Cowie, R. and Tavabi, L. and Schmitt, M. and Alisamir, S. and Amiriparian, S. and Messner, E.-M. and Song, S. and Liu, S. and Zhao, Z. and Mallol-Ragolta, A. and Ren, Z. and Soleymani, M. and Pantic, M.},
  title = {AVEC 2019 workshop and challenge: State-of-mind, detecting depression with AI, and cross-cultural affect recognition},
  booktitle = {Proceedings of the 9th International Workshop on Audio/Visual Emotion Challenge and Workshop, Nice, France},
  year = {2019},
  pages = {3--12},
  doi = {10.1145/3347320.3357688}
}

@inproceedings{gratch2014DAIC,
  author = {Gratch, J. and Artstein, R. and Lucas, G. M. and Stratou, G. and Scherer, S. and Nazarian, A. and Wood, R. and Boberg, J. and DeVault, D. and Marsella, S. and Traum, D. R.},
  title = {The distress analysis interview corpus of human and computer interviews},
  booktitle = {Proceedings of the 9th Language Resources and Evaluation Conference (LREC 2014)},
  year = {2014},
  pages = {3123--3128}
}

@article{yoon2022Dvlog,
  author = {Yoon, J. and Kang, C. and Kim, S. and Han, J.},
  title = {D-vlog: Multimodal vlog dataset for depression detection},
  journal = {Proceedings of the AAAI Conference on Artificial Intelligence},
  year = {2022},
  volume = {36},
  number = {11},
  pages = {12226--12234},
  doi = {10.1609/aaai.v36i11.21483},
  note = {[Data set]}
}

@misc{cai2020MODMA,
  author = {Cai, H. and Gao, Y. and Sun, S. and Li, N. and Tian, F. and Xiao, H. and Li, J. and Yang, Z. and Li, X. and Zhao, Q. and Liu, Z. and Yao, Z. and Yang, M. and Peng, H. and Zhu, J. and Zhang, X. and Hu, X. and Hu, B.},
  title = {MODMA dataset: A multi-modal open dataset for mental-disorder analysis},
  year = {2020},
  howpublished = {arXiv preprint arXiv:2002.09283},
  url = {https://arxiv.org/abs/2002.09283},
  note = {[Data set]}
}

@article{fan2024Transformer,
  author = {Fan, H. and Zhang, X. and Xu, Y. and Fang, J. and Zhang, S. and Zhao, X. and Yu, J.},
  title = {Transformer-based multimodal feature enhancement networks for multimodal depression detection integrating video, audio and remote photoplethysmograph signals},
  journal = {Information Fusion},
  year = {2024},
  volume = {104},
  pages = {102161},
  doi = {10.1016/j.inffus.2023.102161}
}

@article{qin2025Language,
  author = {Qin, R. and Yang, K. and Abbasi, A. and Dobolyi, D. and Seyedi, S. and Griner, E. and Kwon, H. and Cotes, R. and Jiang, Z. and Clifford, G. and Cook, R. A.},
  title = {Language models for online depression detection: A review and benchmark analysis on remote interviews},
  journal = {ACM Transactions on Management Information Systems},
  year = {2025},
  volume = {16},
  number = {2},
  pages = {Article 12},
  doi = {10.1145/3673906}
}

@ARTICLE{Yongfeng2024DepMSTAT,
  author={Tao, Yongfeng and Yang, Minqiang and Li, Huiru and Wu, Yushan and Hu, Bin},
  journal={IEEE Transactions on Knowledge and Data Engineering}, 
  title={DepMSTAT: Multimodal Spatio-Temporal Attentional Transformer for Depression Detection}, 
  year={2024},
  volume={36},
  number={7},
  pages={2956-2966},
  doi={10.1109/TKDE.2024.3350071}
}

@article{Sun2024novel,
  author  = {Sun, C and Jiang, M and Gao, L and Xin, Y and Dong, Y},
  year    = {2024},
  title   = {A novel study for depression detecting using audio signals based on graph neural network},
  journal = {Biomedical Signal Processing and Control},
  volume  = {88},
  pages   = {105675},
  doi     = {10.1016/j.bspc.2023.105675}
}

@article{das2024deep,
  author  = {Das, A. K. and Naskar, R.},
  year    = {2024},
  title   = {A deep learning model for depression detection based on MFCC and CNN generated spectrogram features},
  journal = {Biomedical Signal Processing and Control},
  volume  = {90},
  pages   = {105898},
  doi     = {10.1016/j.bspc.2023.105898}
}

@article{huang2024depression,
  author  = {Huang, X. and Wang, F. and Gao, Y.},
  year    = {2024},
  title   = {Depression recognition using voice-based pre-training model},
  journal = {Scientific Reports},
  volume  = {14},
  pages   = {12734},
  doi     = {10.1038/s41598-024-63556-0}
}

@article{albahrani2021review,
  author  = {Albahrani, E. A. and Alshekly, T. K. and Lafta, S. H.},
  year    = {2021},
  title   = {A review on audio encryption algorithms using chaos maps-based techniques},
  journal = {Journal of Computer Science and Network Security},
  volume  = {11},
  number  = {1},
  pages   = {53--82},
  month   = {11}
}

@article{ustubioglu2022cnt,
  author    = {Ustubioglu, B. and Küçükuğurlu, B. and Ulutas, G.},
  title     = {Robust copy-move detection in digital audio forensics based on pitch and modified discrete cosine transform},
  journal   = {Multimedia Tools and Applications},
  volume    = {81},
  pages     = {27149--27185},
  year      = {2022},
  doi       = {10.1007/s11042-022-13035-3},
  url       = {https://doi.org/10.1007/s11042-022-13035-3}
}

@article{alanazi2023novel,
  author    = {Alanazi, A. S. and Munir, N. and Khan, M. and others},
  title     = {A novel design of audio signals encryption with substitution permutation network based on the Genesio-Tesi chaotic system},
  journal   = {Multimedia Tools and Applications},
  volume    = {82},
  pages     = {26577--26593},
  year      = {2023},
  doi       = {10.1007/s11042-023-14964-3},
  url       = {https://doi.org/10.1007/s11042-023-14964-3}
}

@article{nguyen2024efficient,
  author  = {Nguyen, L. and Phan, B. and Zhang, L. and et al.},
  year    = {2024},
  title   = {An efficient approach for securing audio data in AI training with fully homomorphic encryption},
  journal = {TechRxiv},
  doi     = {10.36227/techrxiv.170956397.78402834/v1},
  note    = {Preprint}
}

@article{william2022switch,
  author  = {William Fedus and Barret Zoph and Noam Shazeer},
  title   = {Switch Transformers: Scaling to Trillion Parameter Models with Simple and Efficient Sparsity},
  journal = {Journal of Machine Learning Research},
  year    = {2022},
  volume  = {23},
  number  = {120},
  pages   = {1--39},
  url     = {http://jmlr.org/papers/v23/21-0998.html}
}

@misc{WHO2023Depression,
  author = {{World Health Organization}},
  organization = {World Health Organization},
  year = {2023},
  month = {March},
  title = {Depressive disorder (depression)},
  howpublished = {\url{https://www.who.int/news-room/fact-sheets/detail/depression}},
  note = {[Accessed: Apr, 29, 2025]}
}

@misc{WHO2025Depression,
  author = {{World Health Organization}},
  organization = {World Health Organization},
  year = {2025},
  title = {Depression Rates by Country 2025},
  howpublished = {\url{https://worldpopulationreview.com/country-rankings/depression-rates-by-country}},
  note = {[Accessed: Apr, 29, 2025]}
}

@article{sardari2022audio,
  author={S. Sardari and B. Nakisa and M. N. Rastgoo and P. Eklund},
  title={Audio based depression detection using Convolutional Autoencoder},
  journal={Expert Systems with Applications},
  year={2022},
  volume={189},
  pages={116076},
  doi={10.1016/j.eswa.2021.116076}
}

@article{Emna2022MFCC,
title = {MFCC-based Recurrent Neural Network for automatic clinical depression recognition and assessment from speech},
journal = {Biomedical Signal Processing and Control},
volume = {71},
pages = {103107},
year = {2022},
issn = {1746-8094},
doi = {https://doi.org/10.1016/j.bspc.2021.103107},
url = {https://www.sciencedirect.com/science/article/pii/S1746809421007047},
author = {Emna Rejaibi and Ali Komaty and Fabrice Meriaudeau and Said Agrebi and Alice Othmani}
}

@article{Xiaolin2022Fusing,
title = {Fusing features of speech for depression classification based on higher-order spectral analysis},
journal = {Speech Communication},
volume = {143},
pages = {46-56},
year = {2022},
issn = {0167-6393},
doi = {https://doi.org/10.1016/j.specom.2022.07.006},
url = {https://www.sciencedirect.com/science/article/pii/S0167639322001029},
author = {Xiaolin Miao and Yao Li and Min Wen and Yongyan Liu and Ibegbu Nnamdi Julian and Hao Guo}
}

@ARTICLE{yan2021Detecting,
  author={Yan ZHAO and Yue XIE and Ruiyu LIANG and Li ZHANG and Li ZHAO and Chengyu LIU },
  journal={IEICE TRANSACTIONS on Information}, 
  title={Detecting Depression from Speech through an Attentive LSTM Network}, 
  year={2021},
  volume={E104-D},
  number={11},
  pages={2019-2023},
  doi={10.1587/transinf.2020EDL8132},
  ISSN={1745-1361},
  month={November},}

@article{Vasa04052023,
author = {Jalpesh Vasa and Amit Thakkar},
title = {Deep Learning: Differential Privacy Preservation in the Era of Big Data},
journal = {Journal of Computer Information Systems},
volume = {63},
number = {3},
pages = {608--631},
year = {2023},
publisher = {Taylor \& Francis},
doi = {10.1080/08874417.2022.2089775},
URL = { 
     https://doi.org/10.1080/08874417.2022.2089775
},
eprint = { 
    https://doi.org/10.1080/08874417.2022.2089775
}
}

@inproceedings {USENIX-audio-1,
author = {Zhiyuan Yu and Yuanhaur Chang and Ning Zhang and Chaowei Xiao},
title = {{SMACK}: Semantically Meaningful Adversarial Audio Attack},
booktitle = {32nd USENIX Security Symposium (USENIX Security 23)},
year = {2023},
isbn = {978-1-939133-37-3},
address = {Anaheim, CA},
pages = {3799--3816},
url = {https://www.usenix.org/conference/usenixsecurity23/presentation/yu-zhiyuan-smack},
publisher = {USENIX Association},
month = aug
}

@INPROCEEDINGS {SP-audio-1,
author = { Lan, Jiahe and Wang, Jie and Yan, Baochen and Yan, Zheng and Bertino, Elisa },
booktitle = { 2024 IEEE Symposium on Security and Privacy (SP) },
title = {{ FlowMur: A Stealthy and Practical Audio Backdoor Attack with Limited Knowledge }},
year = {2024},
volume = {},
ISSN = {},
pages = {1646-1664},
doi = {10.1109/SP54263.2024.00148},
url = {https://doi.ieeecomputersociety.org/10.1109/SP54263.2024.00148},
publisher = {IEEE Computer Society},
address = {Los Alamitos, CA, USA},
month =May}

@ARTICLE{Subspace-Decompo-1,
  author={Deng, Ruixiang and Zhang, Yingwei and Luo, Chaomin and Bi, Zhuming},
  journal={IEEE Transactions on Industrial Informatics}, 
  title={Multimode Process Monitoring Based on Common and Unique Subspace Decomposition}, 
  year={2024},
  volume={20},
  number={9},
  pages={10933-10945},
  doi={10.1109/TII.2024.3397366}}

@article{Subspace-Decompo-2,
title = {Virtual array modeling and performance analysis of half-dimensional real-valued subspace decomposition method},
journal = {Signal Processing},
volume = {240},
pages = {110371},
year = {2026},
issn = {0165-1684},
doi = {https://doi.org/10.1016/j.sigpro.2025.110371},
url = {https://www.sciencedirect.com/science/article/pii/S0165168425004876},
author = {Xiangtian Meng and Bingxia Cao and Lingda Ren and Fenggang Yan and Maria Greco and Fulvio Gini},
}

@INPROCEEDINGS{Subspace-Decompo-3,
  author={Zhang, Gehang and Sheng, Jiawei and Wang, Shicheng and Liu, Tingwen},
  booktitle={ICASSP 2024 - 2024 IEEE International Conference on Acoustics, Speech and Signal Processing (ICASSP)}, 
  title={Noise-Disentangled Graph Contrastive Learning via Low-Rank and Sparse Subspace Decomposition}, 
  year={2024},
  volume={},
  number={},
  pages={5880-5884},
  doi={10.1109/ICASSP48485.2024.10445929}}

@article{Autoencoder-1,
title = {A comprehensive review of synthetic data generation in smart farming by using variational autoencoder and generative adversarial network},
journal = {Engineering Applications of Artificial Intelligence},
volume = {131},
pages = {107881},
year = {2024},
issn = {0952-1976},
doi = {https://doi.org/10.1016/j.engappai.2024.107881},
url = {https://www.sciencedirect.com/science/article/pii/S0952197624000393},
author = {Yaganteeswarudu Akkem and Saroj Kumar Biswas and Aruna Varanasi},
}

@book{spacetime-1,
  title={Spacetime, Geometry, Cosmology},
  author={Burke, W.L.},
  isbn={9780486845586},
  lccn={2020029860},
  url={https://books.google.com.au/books?id=vm8HEAAAQBAJ},
  year={2020},
  publisher={Dover Publications}
}

@article{GAN-1,
author = {Goodfellow, Ian and Pouget-Abadie, Jean and Mirza, Mehdi and Xu, Bing and Warde-Farley, David and Ozair, Sherjil and Courville, Aaron and Bengio, Yoshua},
title = {Generative adversarial networks},
year = {2020},
issue_date = {November 2020},
publisher = {Association for Computing Machinery},
address = {New York, NY, USA},
volume = {63},
number = {11},
issn = {0001-0782},
url = {https://doi.org/10.1145/3422622},
doi = {10.1145/3422622},
journal = {Commun. ACM},
month = oct,
pages = {139–144},
numpages = {6}
}

@article{Differential-Privacy-1,
    author = {Dong, Jinshuo and Roth, Aaron and Su, Weijie J.},
    title = {Gaussian Differential Privacy},
    journal = {Journal of the Royal Statistical Society Series B: Statistical Methodology},
    volume = {84},
    number = {1},
    pages = {3-37},
    year = {2022},
    month = {02},
    issn = {1369-7412},
    doi = {10.1111/rssb.12454},
    url = {https://doi.org/10.1111/rssb.12454},
    eprint = {https://academic.oup.com/jrsssb/article-pdf/84/1/3/49324238/jrsssb_84_1_3.pdf},
}

@article{PHQ-9-1, 
title={Equivalency of the diagnostic accuracy of the PHQ-8 and PHQ-9: a systematic review and individual participant data meta-analysis}, 
volume={50}, 
DOI={10.1017/S0033291719001314}, 
number={8}, 
journal={Psychological Medicine}, 
author={Wu, Yin and Levis, Brooke and Riehm, Kira E. and Saadat, Nazanin and Levis, Alexander W. and Azar, Marleine and Rice, Danielle B. and Boruff, Jill and Cuijpers, Pim and Gilbody, Simon and et al.}, 
year={2020}, 
pages={1368–1380}}

@article{VoicePrint-1,
  author       = {Kroenke, Kurt},
  title        = {PHQ-9: Global uptake of a depression scale},
  journal      = {World Psychiatry},
  year         = {2021},
  volume       = {20},
  number       = {1},
  pages        = {135--136},
  doi          = {10.1002/wps.20821},
  url          = {https://doi.org/10.1002/wps.20821}
}

@inproceedings{CCS-1,
  author = {Wang, Penghao and Huai, Shuo and Cao, Yetong and Liu, Chao and Luo, Jun},
  title = {Threat from Windshield: Vehicle Windows as Involuntary Attack Sources on Automotive Voice Assistants},
  booktitle = {Proceedings of the ACM Conference on Computer and Communications Security (CCS)},
  year = {2025},
  month = {10},
  publisher = {ACM},
  doi = {10.1145/3719027.3765171}
}

@inproceedings{NDSS-1,
  author    = {Guangke Chen and Yedi Zhang and Fu Song and Ting Wang and Xiaoning Du and Yang Liu},
  title     = {SongBsAb: A Dual Prevention Approach against Singing Voice Conversion based Illegal Song Covers},
  booktitle = {Proceedings of the Network and Distributed System Security Symposium (NDSS)},
  year      = {2025},
  organization = {NDSS},
}

@inproceedings {USENIX-2,
author = {Jiangyi Deng and Fei Teng and Yanjiao Chen and Xiaofu Chen and Zhaohui Wang and Wenyuan Xu},
title = {{V-Cloak}: Intelligibility-, Naturalness- \& {Timbre-Preserving} {Real-Time} Voice Anonymization},
booktitle = {32nd USENIX Security Symposium (USENIX Security 23)},
year = {2023},
isbn = {978-1-939133-37-3},
address = {Anaheim, CA},
pages = {5181--5198},
url = {https://www.usenix.org/conference/usenixsecurity23/presentation/deng-jiangyi-v-cloak},
publisher = {USENIX Association},
month = aug
}

@INPROCEEDINGS{SP-2,
  author={Wang, Haodi and Dong, Kai and Zhu, Zhilei and Qin, Haotong and Liu, Aishan and Fang, Xiaolin and Wang, Jiakai and Liu, Xianglong},
  booktitle={2024 IEEE Symposium on Security and Privacy (SP)}, 
  title={Transferable Multimodal Attack on Vision-Language Pre-training Models}, 
  year={2024},
  volume={},
  number={},
  pages={1722-1740},
  doi={10.1109/SP54263.2024.00102}}

@inproceedings{CCS-2,
  author    = {Luming Yang and Lin Liu and Jun-Jie Huang and Zhuotao Liu and Shiyu Liang and Shaojing Fu and Yongjun Wang},
  title     = {MM4flow: A Pre-trained Multi-modal Model for Versatile Network Traffic Analysis},
  booktitle = {Proceedings of the 32nd ACM Conference on Computer and Communications Security (CCS)},
  year      = {2025},
  organization = {ACM},
}

@inproceedings {USENIX-3,
author = {Kaiming Cheng and Jeffery F. Tian and Tadayoshi Kohno and Franziska Roesner},
title = {Exploring User Reactions and Mental Models Towards Perceptual Manipulation Attacks in Mixed Reality},
booktitle = {32nd USENIX Security Symposium (USENIX Security 23)},
year = {2023},
isbn = {978-1-939133-37-3},
address = {Anaheim, CA},
pages = {911--928},
url = {https://www.usenix.org/conference/usenixsecurity23/presentation/cheng-kaiming},
publisher = {USENIX Association},
month = aug
}

@inproceedings{NDSS-2,
  author    = {Shu Wang and Kun Sun and Qi Li},
  title     = {Compensating Removed Frequency Components: Thwarting Voice Spectrum Reduction Attacks},
  booktitle = {Proceedings of the Network and Distributed System Security Symposium (NDSS)},
  year      = {2024},
  organization = {NDSS},
}

@INPROCEEDINGS {SP-3,
author = { Murakami, Takao and Sei, Yuichi and Eriguchi, Reo },
booktitle = { 2025 IEEE Symposium on Security and Privacy (SP) },
title = {{ Augmented Shuffle Protocols for Accurate and Robust Frequency Estimation Under Differential Privacy }},
year = {2025},
volume = {},
ISSN = {},
pages = {3892-3911},
doi = {10.1109/SP61157.2025.00019},
url = {https://doi.ieeecomputersociety.org/10.1109/SP61157.2025.00019},
publisher = {IEEE Computer Society},
address = {Los Alamitos, CA, USA},
month =May}

@INPROCEEDINGS {SP-4,
author = { Lou, Jiadong and Yuan, Xu and Zhang, Rui and Yuan, Xingliang and Gong, Neil Zhenqiang and Tzeng, Nian-Feng },
booktitle = { 2025 IEEE Symposium on Security and Privacy (SP) },
title = {{ GRID: Protecting Training Graph from Link Stealing Attacks on GNN Models }},
year = {2025},
volume = {},
ISSN = {},
pages = {2095-2113},
doi = {10.1109/SP61157.2025.00059},
url = {https://doi.ieeecomputersociety.org/10.1109/SP61157.2025.00059},
publisher = {IEEE Computer Society},
address = {Los Alamitos, CA, USA},
month =May}

@INPROCEEDINGS{SP-5,
  author={David, Liron and Berkman, Omer and Hassidim, Avinatan and Lazarov, David and Matias, Yossi and Yung, Moti},
  booktitle={2025 IEEE Symposium on Security and Privacy (SP)}, 
  title={Extended Diffie-Hellman Encryption for Secure and Efficient Real-Time Beacon Notifications}, 
  year={2025},
  volume={},
  number={},
  pages={4406-4418},
  doi={10.1109/SP61157.2025.00230}}
